\documentclass[twocolumn,showpacs,preprintnumbers,amsmath,amssymb]{revtex4}
\usepackage{graphicx}
\usepackage{dcolumn}
\usepackage{bm}
\begin{document}
\title{Muon-Induced Background Study for Underground Laboratories }
\author{
D.-M. Mei and A. Hime 
}
\affiliation{
Physics Division, MS H803, Los Alamos National Laboratory, Los Alamos, NM 87545, USA\\
}%
\date{December 5, 2005}

\begin{abstract}
We provide a comprehensive study of the cosmic-ray muon flux and induced activity as a function of overburden along with a convenient parameterization of the salient fluxes and differential distributions for a suite of underground laboratories ranging in depth from $\sim$1 to 8 km.w.e.. Particular attention is given to the muon-induced fast neutron activity for the underground sites and we develop a Depth-Sensitivity-Relation to characterize the effect of such background in experiments searching for WIMP dark matter and neutrinoless double beta decay.
\end{abstract}

\maketitle
\section{Introduction}
Underground laboratories provide the overburden necessary for experiments sensitive to cosmic-ray muons and their progenies. Muons traversing a detector and its surrounding material that miss an external veto serve as a background themselves and secondary backgrounds are induced in the production of fast neutrons and cosmogenic radioactivity. 
In this study we have focused on the muon-induced fast neutron background as a function of depth and the implications for rare event searches for neutrinoless double beta decay and WIMP dark matter. One of our main goals is to develop a Depth-Sensitivity-Relation (DSR) in terms of the total muon and muon-induced neutron flux and to put this into the context of existing underground laboratories covering a wide range of overburden.

In Section II we review the experimental data available for differential muon fluxes and provide a definition of depth in terms of the total muon flux that removes some confusion regarding the equivalent depth of an underground site situated under a mountain versus one with flat overburden. The muon fluxes and differential distributions are parameterized and used as input in Section III to generate, via FLUKA simulations~\cite{fluka}, the production rate for fast neutrons. The total neutron flux and salient distributions are compared with the available experimental data and we provide some convenient parameterizations that can be used as input for detector-specific simulations at a given underground site. We quantify the agreement between FLUKA simulation and experimental data and provide an explanation for the discrepancy between neutron flux and energy spectra as measured in the LVD detector. Muon-induced cosmogenic radioactivity is discussed in terms of depth and the average muon energy in Section IV. In Section V we apply our results to a generic study of germanium-based experiments in search of neutrinoless double beta decay and WIMP dark matter and demonstrate the utility of the DSR in projecting the sensitivity and depth requirements of such experiments. We conclude with a summary of the results and an outline of new studies underway.

\section{Depth-Intensity-Relation and Distributions for Cosmic-Ray Muons}
\subsection{Through-Going Muon Intensity}
\subsubsection{Differential Muon Intensity versus Slant-Depth}
\begin{figure}[htb!!!]
\includegraphics[angle=0,width=8.6cm] {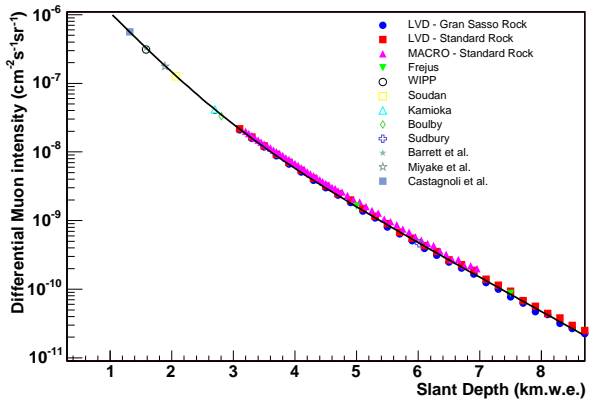}
\caption{\small{
Measurements of the differential muon flux as a function of slant depth from Castagnoli~\cite{cca}, Barrett~\cite{phb}, Miyake~\cite{smi}, WIPP~\cite{eie}, Soudan~\cite{kasa}, Kamioka~\cite{kamland}, 
Boulby~\cite{mro}, Gran Sasso~\cite{lvdd,macroo}, Fr\'{e}jus~\cite{chb} and Sudbury~\cite{andrew}. 
Note that the measurements for Kamioka~\cite{kamland} and Sudbury~\cite{andrew} 
are reported as the number of muons per day. We calculate the effective detector acceptance for these two measurements in order to obtain the muon flux. The solid curve is our global fit function described by equation~(\ref{eq:muon1}).}}
\label{fig:intensity}
\end{figure}

The cosmic-ray muon flux in the atmosphere, underground, and underwater has been a subject of study for more than five decades ~\cite{evbu}. Experimental data on the differential muon intensity (in units of $cm^{-2}s^{-1}sr^{-1}$) are shown in Fig.~\ref{fig:intensity} as a function of slant-depth measured in units of kilometers of water equivalent (km.w.e.), where 1000 hg/cm$^{2}$ = 10$^{5}$ g/cm$^{2}$ $\equiv$ 1 km.w.e..

Groom $et.$ $al.$ proposed a model~\cite{glc} to fit the experimental data to a Depth-Intensity-Relation (DIR), appropriate for the range (1 - 10 km.w.e.):
\begin{equation}
\label{eq:muon1}
I(h)=(I_{1}e^{(-h/\lambda_{1})}+I_{2}e^{(-h/\lambda_{2})}),
\end{equation}
where I$(h)$ is the differential muon intensity corresponding to the slant-depth, $h$.

\begin{figure}[htb!!!]
\includegraphics[angle=0,width=8.6cm] {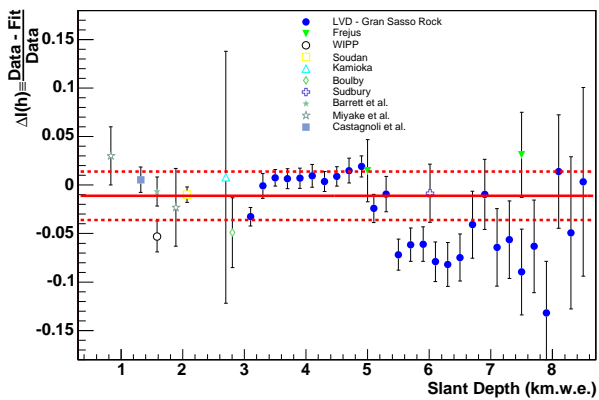}
\caption{\small{
The relative deviation between the global fit function and the measured data on the differential muon flux from Castagnoli~\cite{cca}, 
Barrett~\cite{phb},
Miyake~\cite{smi}, WIPP~\cite{eie}, Soudan~\cite{kasa}, Kamioka~\cite{kamland}, 
Boulby~\cite{mro}, Gran Sasso~\cite{lvdd}, Fr\'{e}jus~\cite{chb} and Sudbury~\cite{andrew}. The horizontal lines indicate the root-mean-square deviation amongst the residuals.}}
\label{fig:muonderivation}
\end{figure}

Using the experimental data in Fig.~\ref{fig:intensity} we determine the free parameters of equation~(\ref{eq:muon1}) as: 
I$_{1}$ = (8.60$\pm0.53)\times$ 10$^{-6}$ 
sec$^{-1}$cm$^{-2}$sr$^{-1}$, I$_{2}$ = (0.44$\pm0.06)\times$ 10$^{-6}$ 
sec$^{-1}$cm$^{-2}$sr$^{-1}$, $\lambda_{1}$ = 0.45$\pm$0.01 km.w.e., $\lambda_{2}$ = 0.87$\pm$0.02
km.w.e.. The relative deviation between the data and our fit is shown in Fig.~\ref{fig:muonderivation}, indicating that the parameterization reproduces the experimental data reasonably well and with an overall accuracy of about 5\%.

\subsubsection{The Total Muon Flux with Flat Overburden}
For an underground laboratory with flat overburden it is straightforward to calculate the total muon intensity arriving below the surface at a vertical depth, $h_{0}$. In the flat earth approximation, the through-going muon intensity (I$_{th}$)
 for a specific slant-depth, $h$, in the direction of zenith angle, $\theta$, reads:
\begin{equation}
\label{eq:muon2}
I_{th}(h,\theta) = I(h)G(h,\theta),
\end{equation}
where G($h$,$\theta$) = sec($\theta$), $h$  = $h_{0}$sec($\theta$), and $I(h)$ is the DIR expressed in  equation~(\ref{eq:muon1}). Equation~(\ref{eq:muon1}) now becomes
\begin{equation}
\label{eq:muon3}
I_{th}(h,\theta) = (I_{1}e^{(-h_{0}sec(\theta)/\lambda_{1})}+I_{2}e^{(-h_{0}sec(\theta)/\lambda_{2})})sec(\theta). 
\end{equation}
Integration over the 
upper hemisphere using  equation (\ref{eq:muon3}) then provides the total muon intensity for an underground site with flat overburden  positioned at a vertical depth $h_{0}$.

Using the experimental data for the total muon flux and knowledge of the vertical depth for a set of underground sites with flat overburden (WIPP~\cite{eie}, Soudan~\cite{kama},
Boulby~\cite{mro} and Sudbury~\cite{andrew}) we can now define a fit-function which is similar to the differential muon intensity function (equation~(\ref{eq:muon1})):
\begin{equation}
\label{eq:intergalgroom}
I_{\mu}(h_{0}) = 67.97 \times 10^{-6}e^{\frac{-h_{0}}{0.285}}+2.071 \times 10^{-6}e^{\frac{-h_{0}}{0.698}},
\end{equation}
where $h_{0}$ is the vertical depth in km.w.e. and I$_{\mu}(h_{0}$) is in units of cm$^{-2}s^{-1}$, appropriate in the flat-earth approximation.
 
\subsubsection{The Total Muon Flux in Case of Mountain Overburden}
In the case that a laboratory is situated underneath a mountain, additional information regarding the mountain shape or elevation map, h($\theta,\phi$), is required to determine the total muon flux:
\begin{equation}
\label{eq:mount}
I_{tot} = \int sin(\theta) d\theta \int d\phi I(h(\theta,\phi)) G(h,\theta),
\end{equation}
where G($h$,$\theta$) = sec($\theta$) and $I_{tot}$ is the total muon flux obtained after integrating over the mountain shape and using the DIR defined in equation~(\ref{eq:muon1}).

As an example, we have computed the total muon flux at the Gran Sasso Laboratory using the detailed information provided by the MACRO collaboration~\cite{macroo} on the mountain shape and their measurements of the differential muon flux (see Fig.~\ref{fig:intensity}). We find a total muon intensity of (2.58$\pm0.3)\times10^{-8} cm^{-2}sec^{-1}$, which is consistent within about 20\% to that obtained in Refs.~\cite{hwuu,gallex}. If this intensity is now entered into the left hand side of equation (\ref{eq:intergalgroom}), we can now solve for the equivalent vertical depth relative to a flat overburden for the Gran Sasso Laboratory and find it to be 3.1 $\pm$ 0.2 km.w.e.. 

This depth should not be confused with the average depth that would be deduced simply by integrating over the depth profile of the mountain:
\begin{equation}
\label{eq:avgdepth}
<h> = \int sin(\theta) d\theta \int d\phi h(\theta,\phi),
\end{equation}
which yields 3.65 km.w.e.. A similar approach can be taken with information available from the Fr\'{e}jus Collaboration~\cite{chb}. We find a total muon intensity of (4.83$\pm0.5)\times10^{-9} cm^{-2}sec^{-1}$ corresponding to an equivalent flat overburden of 4.2 $\pm$ 0.2 km.w.e. and an average depth of 5 km.w.e.. Our calculation is consistent with the Fr\'{e}jus Collaboration's result within 12\%. We note that the equivalent ``flat-overburden'' depth defined by the experimental measure of the total muon flux is $\sim$15-20\% lower than that often quoted for Gran Sasso and Frejus based on the average physical depth.

\subsubsection{Definition of Depth and Total Muon Flux for Underground Sites} 

\begin{figure}[htb!!!]
\includegraphics[angle=0,width=8.6cm]{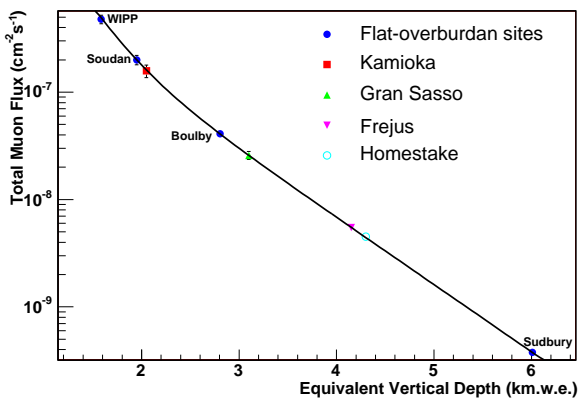}
\caption{\small{
The total muon flux measured for the various underground sites summarized in Table~\ref{tab:muonflux2} as a function of the equivalent vertical depth relative to a flat overburden. The smooth curve is our global fit function to those data taken from sites with flat overburden (equation~(\ref{eq:intergalgroom})).}}
\label{fig:integral1}
\end{figure}

\begin{figure}[htb!!!]
\includegraphics[angle=0,width=8.6cm]{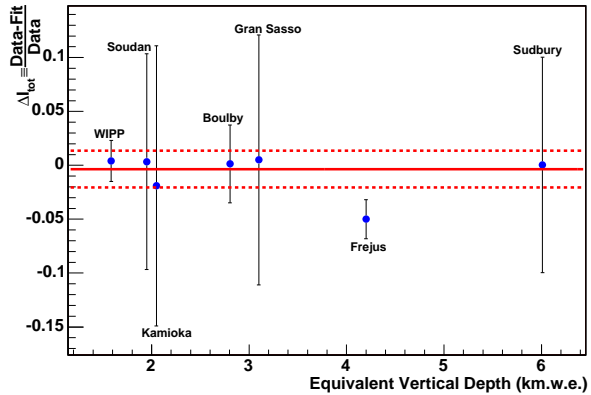}
\caption{\small{
The relative deviation between data on the total muon flux and our global fit function. The horizontal lines indicate the root-mean-square deviation amongst the residuals based upon the experimental uncertainties in the measurements.}}
\label{fig:totalresidual}
\end{figure}

The data on the total muon intensity at the various underground sites is summarized in Table~\ref{tab:muonflux2} and Fig.~\ref{fig:integral1}. We use equation~(\ref{eq:intergalgroom}) to calculate the total muon flux for Homestake (flat-overburden) at the depth 4.3 $\pm$ 0.2 km.w.e.~\cite{home}.  The relative difference between the data and our model (equation~(\ref{eq:intergalgroom})) is shown in Fig.~\ref{fig:totalresidual}, where the uncertainties reflect the experimental uncertainties in Table~\ref{tab:muonflux2}. In order to circumvent the misuse of vertical muon intensity in comparing sites with flat overburden to those under mountains, we define the equivalent depth relative to a flat overburden by the experimental measurements of the total muon intensity. This definition and these intensities are used hereafter.
 
\begin{table}[htb!!!]
\caption{Summary of the total muon flux measured at the underground sites and the equivalent vertical depth relative to a flat overburden.}
\label{tab:muonflux2}
\begin{tabular}{lll}
\hline \hline
Site  &Total flux & Depth\\
      &cm$^{-2}$sec$^{-1}$& km.w.e.\\
\hline
WIPP  &(4.77$\pm0.09)\times10^{-7}$~\cite{eie}&1.585$\pm$0.011 \\
Soudan &(2.0$\pm0.2)\times10^{-7}$~\cite{kama}&1.95$\pm$0.15\\
Kamioka&  (1.58$\pm0.21)\times10^{-7}$~\cite{kamland}&2.05$\pm$0.15$^\dagger$\\
Boulby &(4.09$\pm0.15)\times10^{-8}$~\cite{mro}&2.805$\pm$0.015\\
Gran Sasso & (2.58$\pm0.3)\times10^{-8}$[this work] &3.1$\pm$0.2$^\dagger$ \\
           & (2.78$\pm0.2)\times10^{-8}$~\cite{hwuu}&3.05$\pm$0.2$^\dagger$\\
           & (3.22$\pm0.2)\times10^{-8}$~\cite{gallex}&2.96$\pm$0.2$^\dagger$ \\
Fr\'{e}jus  &(5.47$\pm0.1)\times10^{-9}$~\cite{chb}& 4.15$\pm$0.2$^\dagger$\\
        &(4.83 $\pm0.5)\times10^{-9}$ [this work]&4.2$\pm$0.2$^\dagger$\\
Homestake &(4.4$\pm0.1\times10^{-9})$[this work]&4.3$\pm$0.2\\
Sudbury&(3.77$\pm0.41)\times10^{-10}$~\cite{andrew}&6.011$\pm$0.1 \\
\hline \hline
\end{tabular}

$^\dagger$ Equivalent vertical depth with a flat overburden  
determined by the measured total muon flux.
\end{table}

\subsection{Stopping Muon Intensity}
 Stopping-muons are also a source of background.  For
example, $\mu^{-}$ capture on a nucleus produces neutrons and
radioactive isotopes.  
The total stopping-muon rate has contributions from cosmic-ray muons coming to the end of their range, secondary muons  generated locally through
interactions of the primary muons (due to virtual-photo interactions with nuclei), and local muon production by real photons ($\pi_{0}$-decay in 
electromagnetic showers).  It is customary to quote results in terms 
of the ratio, R, of stopping muons to through-going muons. A detailed
 calculation is provided by Cassiday et al.~\cite{glc}. 
The total ratio, R$(h)$, of stopping-muons to through-going muons (vertical
 direction) at different depths can be parameterized
 as~\cite{tkg}
\begin{equation}
\label{eq:muon4}
R(h) \approx \gamma_{\mu}\frac{\Delta E e^{h/\xi}}{(e^{h/\xi} -1)\epsilon_{\mu}},
\end{equation}
where $\gamma_{\mu}$ = 3.77 for E$_{\mu} \geq$ 1000 GeV~\cite{plt}, $\xi$ =
2.5 km.w.e.,
$\Delta$E $\approx \alpha h$, $\alpha$ = 0.268  
GeV/km.w.e.~\cite{rpp} for   E$_{\mu} \geq$ 1000 GeV~\cite{plt}, 
$h$ is the depth of an underground laboratory, and 
{$\epsilon_{\mu}$ = 618 GeV~\cite{plt}. For large depths, as can be seen 
in Fig.~\ref{fig:stopmuon}, this ratio is
less than 0.5\% and is hereafter neglected for the underground sites considered in this study.

\begin{figure}[htb!!!]
\includegraphics[angle=0,width=8.6cm] {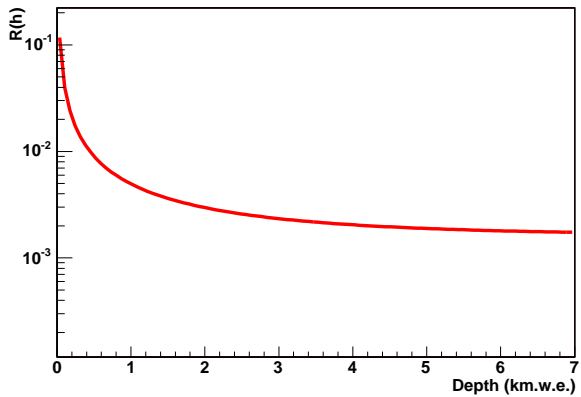}
\caption{\small{
The ratio of stopping-muons to through-going muons, relative to the vertical direction, as a function of depth.}}
\label{fig:stopmuon}
\end{figure}

\subsection{Muon Energy Spectrum and Angular Distribution}
In addition to the total muon intensity arriving at a given underground site, we require knowledge of the differential energy and angular distributions in order to generate the muon-induced activity within a particular experimental cavern. The energy spectrum is discussed 
in Refs.~\cite{tkg,pdg}:
\begin{equation}
\label{eq:musurface5}
\frac{dN}{dE_{\mu}} = A e^{-bh(\gamma_{\mu}-1)}\cdot (E_{\mu} + \epsilon_{\mu} (1-e^{-bh}))^{-\gamma_{\mu}},
\end{equation}
where A is a normalization constant with respect to the differential muon intensity
at a given depth and E$_{\mu}$ is the muon energy after crossing
the rock slant depth $h$ (km.w.e.). Fig.~\ref{fig:muenspe} shows the local muon energy spectrum for the various underground laboratories under consideration using the parameters
b = 0.4/km.w.e., $\gamma_{\mu}$ = 3.77 and $\epsilon_{\mu}$ =  693 GeV~\cite{deg}.  
Fig.~\ref{fig:muenang} shows the local angular distribution for the same sites where we assume a sec($\theta)$ distribution, valid for depths in excess of 1.5 km.w.e.~\cite{mgk}. 
Note that the overall angular distribution of muons at the surface is proportional to
cos$^{2}(\theta)$ with an average muon energy of about 4 GeV~\cite{pdg}.
 
\begin{figure}[htb!!!]
\includegraphics[angle=0,width=8.6cm] {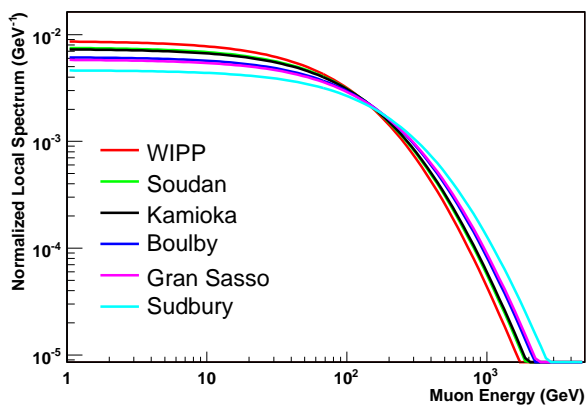}
\caption{\small{
The muon energy spectrum local to the various underground sites calculated using equation (\ref{eq:musurface5}). The areas 
under the curves are normalized to the vertical muon
 intensity for comparison purposes.}}
\label{fig:muenspe}
\end{figure}

\begin{figure}[htb!!!]
\includegraphics[angle=0,width=8.6cm] {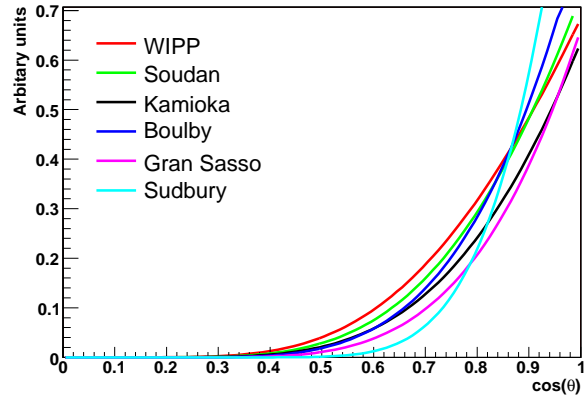}
\caption{\small{
The muon angular distribution local to the various underground sites
based on equation (\ref{eq:muon3}). All curves have been normalized to the total muon intensity for comparison purposes.}}
\label{fig:muenang}
\end{figure}

From equation (\ref{eq:musurface5}), the average muon energy at depth $h$ is given by:
\begin{equation}
\label{eq:musurface6}
<E_{\mu}> = \frac{\epsilon_{\mu}(1-e^{-bh})}{\gamma_{\mu} -2}.
\end{equation}
The parameters $\epsilon_{\mu}$, b and $\gamma_{\mu}$ in equation~(\ref{eq:musurface6}) have been studied
by several authors~\cite{plt,deg,gba} for standard rock 
(A = 22, Z = 11, $\rho$ = 2.65 g cm$^{-2}$). Uncertainty in these
parameters are due to uncertainties in the muon energy spectrum in the atmosphere, details of muon energy loss in the media, and 
the local rock density and composition. Table~\ref{tab:muonenergy1} summarizes the average muon energy
for the various sites where we have used two different sets of parameters provided by Lipari et al. ( b = 0.383/km.w.e., $\gamma_{\mu}$ = 3.7 and 
$\epsilon_{\mu}$ =  618 GeV~\cite{plt}) and Groom et al. (b = 0.4/km.w.e.~\cite{gba}, $\gamma_{\mu}$ = 3.77 and $\epsilon_{\mu}$ = 693 GeV~\cite{deg}). 
The measured average single muon energy at Gran Sasso~\cite{mam} is 270$\pm 3(stat) \pm 18(syst)$ GeV 
which has an uncertainty of 6.8\%. The predicted values using both
 sets of parameters agree with the measured value within the measured  uncertainty. 

\begin{table}[htb!!!]
\caption{Single muon average energies for the various underground sites.}
\label{tab:muonenergy1}
\begin{tabular}{llll}
\hline \hline
Site & Lipari et al. & Groom et al. & Measured value\\
\hline
WIPP  & 165 GeV & 184 GeV & \\
Soudan& 191 GeV & 212 GeV & \\
Kamioka& 198 GeV & 219 GeV & \\
Boulby & 239 GeV& 264 GeV & \\
Gran Sasso & 253  GeV& 278 GeV & 270$\pm$18 GeV~\cite{mam}\\
Sudbury& 327 GeV& 356 GeV & \\
\hline \hline
\end{tabular}
\end{table}

\section{\label{scuts}Muon-Induced Neutrons}
We distinguish two classes
of fast neutrons, namely neutrons produced by muons traversing the detector
itself, and neutrons created in the external rock by muons missing the
veto detector. The former can be tagged effectively in an external veto with sufficient efficiency surrounding a central detector. The latter are more difficult to shield or tag in coincidence with the primary muon owing to the hard energy spectrum and long propagation range. Thus, we focus here on the fast neutrons produced in the external rock and quantify the production rate as a function of depth.

The production of fast neutrons depends strongly on the depth and composition of an underground site. Generally speaking, the neutron production rate at large depths due to muons is two to three orders of magnitude smaller than that of neutrons arising from local radioactivity through ($\alpha$,n) reactions. Nonetheless, the latter process is common to any underground experiment and the low energy neutrons (typically $<$8 MeV) produced via ($\alpha$,n) reactions are relatively easy to shield (see Section V.D). The muon-induced neutrons, on the other hand, have a very hard energy spectrum (extending to several GeV) and can penetrate to significant depth both in the surrounding rock and detector shielding materials. In this section we exploit the total muon fluxes and distributions developed in the previous section as input to FLUKA  simulations to study the muon-induced
neutron flux, energy spectrum, angular distribution, multiplicity, and
lateral distribution in the underground laboratories.

Table~\ref{tab:rockcom} exhibits the rock composition for the six sites under consideration.
The rock composition of
the WIPP site is mainly NaCl~\cite{carlsbad}, whereas for the other four sites, 
the average atomic weight and average atomic number are calculated
based on the known local rock composition~\cite{kasa, mro, macroo, sno}. For lack of additional information we assume standard rock for Kamioka. 
Note that Boulby is a salt mine but the rock composition provided in
Ref.~\cite{macroo} is very similar to the standard rock.

\begin{table}[htb!!!]
\caption{Average matter properties of the various underground sites.}
\label{tab:rockcom}
\begin{tabular}{lllll}
\hline \hline
Site & $<A>$ & $<Z>$ &$<Z>$/$<A>$& g/cm$^{3}$ \\
\hline
WIPP  & 30.0 & 14.64 &0.488& 2.3\\
Soudan& 24.47 & 12.15 &0.497& 2.8\\
Kamioka& 22.0 & 11.0 & 0.5&2.65\\
Boulby & 23.6 & 11.7 &0.496 &2.7 \\
Gran Sasso & 22.87 & 11.41 &0.499& 2.71\\
Sudbury& 24.77 & 12.15 & 0.491&2.894\\
\hline \hline
\end{tabular}
\end{table}

The rock thickness employed in the simulation is 20 m $\times$ 20 m $\times$ 20 m. The laboratory cavern of size
6 m $\times$ 6 m $\times$ 6 m was placed inside the rock region at a depth of 7 m from 
each side of the cube.
This cube ensures equilibrium between neutron and muon fluxes, hence the ratio of neutron to muon fluxes is constant.
 
\subsection{\label{scuts}Comparison Between Data and Simulation}
Table~\ref{tab:rate} lists the mean neutron production rates from seven measurements ~\cite{rht,fbh,lbb,rie,mag,mal} using liquid scintillator covering a significant range in depth and mean muon energy. We provide a global fit function to the data as a function of mean muon energy (see Fig.~\ref{fig:muonenergy1}) and compare this to the Monte Carlo calculations performed in Ref.~\cite{yfg} (C$_{10}$H$_{22}$), Ref.~\cite{vak} (C$_{10}$H$_{20}$) and our FLUKA simulation (C$_{10}$H$_{20}$). For experiments which did not provide the mean muon energy, we use
the experimental depth and the muon energy loss rate~\cite{rpp} to estimate
the mean muon energy.
 
\begin{table}[htb!!!]
\caption{Measured neutron production rates.}
\label{tab:rate}
\begin{tabular}{l|l|l|l}
\hline \hline
 Measurements&Depth & $<E_{\mu}>$ & $<n>$\\ 
 &     km.w.e. & GeV         &   n/$(\mu  \; g \; cm^{-2})$\\
\hline
Hertenberger~\cite{rht} &0.02& 13  & (2$\pm0.7)\times10^{-5}$ \\
Bezrukov~\cite{lbb}& 0.025&14.7 & (4.7$\pm0.5)\times10^{-5}$ \\
Boehm~\cite{fbh}& 0.032&16.5 & (3.6$\pm0.31)\times10^{-5}$ \\
Bezrukov~\cite{lbb}&0.316& 55 & (1.21$\pm0.12)\times10^{-4}$ \\
Enikeev~\cite{rie} &0.75& 120 & (2.15$\pm0.15)\times10^{-4}$ \\
the LVD data~\cite{mal}&3.1 & 270 & (1.5$\pm0.4)\times10^{-4}$\\
Aglietta (the LSD)~\cite{mag}&5.0& 346 & (5.3$\pm1.1)\times10^{-4}$ \\
\hline \hline
\end{tabular}
\end{table}
 
Note that the LVD (see Fig.~\ref{fig:muonenergy1} and Table~\ref{tab:rate}) result obtained at Gran Sasso deviates significantly from the global-fit curve and simulations. Not only is the measured flux apparently low, the differential energy spectrum of fast neutrons measured in the same experiment is also inconsistent with simulation. Kudryavtsev et al.~\cite{vak} suggested that the quenching of proton-recoil energy in the liquid scintillator of LVD~\cite{mal} is a natural explanation for the discrepancy between simulation and the measured energy spectrum. We propose that this same effect is also responsible for the discrepancy in the measured neutron flux.

Following directly from Ref.~\cite{mal} describing the LVD analysis for neutrons, ``a high-energy-threshold (HET) trigger is set at 4-5 MeV. During the 1 ms time period following an HET trigger, a low-energy-threshold (LET) is enabled for counters belonging to the same quarter of the tower which allows the detection of the 2.2 MeV photons from neutron capture by protons. Each neutron ideally should generate two pulses: the first pulse above the HET is due to the recoil protons from n-p elastic scattering (its amplitude is proportional to and even close to the neutron energy); the second pulse, above the LET in the time gate of about 1 ms is due to the 2.2 MeV gamma from neutron capture by a proton. The sequence of two pulses (one above HET and one above LET) was the signature for neutron detection. The energy of the first pulse (above HET) was measured and attributed to the neutron energy.''

The authors go on to ``Note that really this is not the neutron energy but the energy transfered to protons in the scintillator and measured by the counter'', however, they do not correct the visible energy for quenching effects to yield the true proton-recoil energy. Due to the finite HET, this also means that the total number of neutrons counted is also underestimated.
Quenching of protons in scintillator was measured by Ref.~\cite{rac}
and a factor of 2.15 is expected for 4 MeV energy of proton recoil as shown in Fig.~\ref{fig:quen}. If we correct the LVD results by taking this quenching factor and energy threshold into account the total neutron production rate we obtain is 4.5$\times10^{-4}$ n/$(\mu  \; g \; cm^{-2})$, an increase of about a factor of three from the published value~\cite{mal} and consistent with our global fit curve.

One should clarify that such a correction does not apply to the LSD data~\cite{mag} taken from an experiment similar to LVD and operated at Mont Blanc Laboratory. In LSD, no attempt was made to measure the energy of the muon-induced neutrons, however, neutrons were counted by demanding a HET (25-30 MeV) produced by muons with a track length of at least 15 cm in the liquid scintillator. Neutrons were then tagged in coincidence with the delayed capture gamma-ray. Consequently, apart from minor corrections owing to those initial muons producing coincident neutrons that miss the muon trigger, there should be no significant threshold correction associated with neutron counting in LSD.

\begin{figure}[htb!!!]
\includegraphics[angle=0,width=8.6cm]{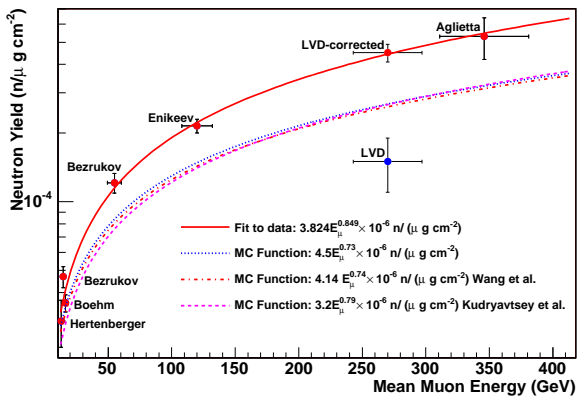}
\caption{\small{
The neutron production rate in liquid scintillator versus the mean muon energy. Data points with uncertainties are experimental measurements from Hertenberger~\cite{rht}, Boehm~\cite{fbh}, Bezrukov~\cite{lbb}, Enikeev~\cite{rie}, the LVD
data~\cite{mal} and Aglietta~\cite{mag}. The solid curve is our global fit to the data after correcting the LVD data point for quenching effects described in the text. Our global fit curve describes the data well but the FLUKA simulations tend to underestimate the neutron production rate by about 35\%.}}
\label{fig:muonenergy1}
\end{figure}

\begin{figure}[htb!!!]
\includegraphics[angle=0,width=8.6cm]{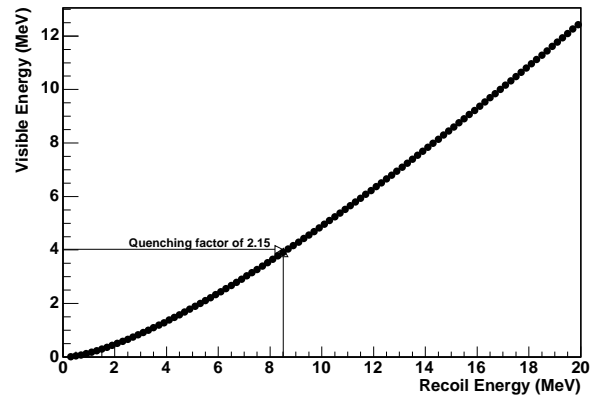}
\caption{\small{The quenching function relating the effective or visible energy in liquid scintillator versus the kinetic energy imparted to a recoiling proton induced by neutron scattering (adopted from Ref.~\cite{rac}).}}
\label{fig:quen}
\end{figure}

 \begin{figure}[htb!!!]
\includegraphics[angle=0,width=8.6cm]{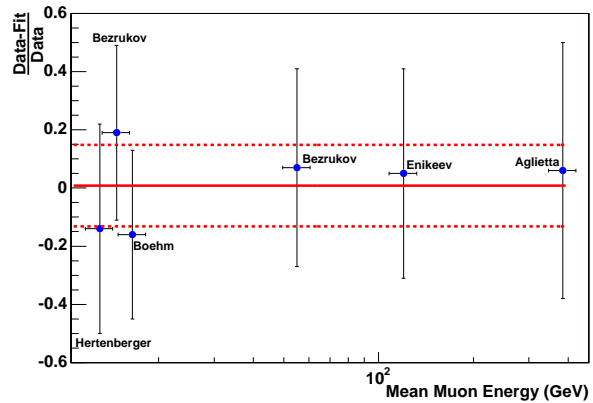}
\caption{\small{The relative deviation between our global fit function and the measured neutron production rate as a function of the mean muon energy. The uncertainties represent the experimental uncertainties on the measured data points and the horizontal lines indicate the root-mean-square deviation on the residuals.}}
\label{fig:neuproderror}
\end{figure} 

As can be seen in Fig's~\ref{fig:muonenergy1} and ~\ref{fig:neuproderror}, the data are well described by a simple power law model suggested by Refs.~\cite{yfg, vak, afk} and our FLUKA simulation. The FLUKA simulations, however, underestimate the data (and the simple power law fit to this data) by about 30 percent. It is natural to attribute this to the virtual photonuclear cross section which is not well known for high energy cosmic-ray muon interactions with nuclei. Nonetheless, the 
integrated cross section of virtual photonuclear interactions of muons measured by
MACRO ~\cite{gba} and ATLAS ~\cite{atlas} show agreement with the Bezrukov-Bugaev model~\cite{bebu} used in FLUKA, though the accuracy of the prediction is limited by the lack of data for muon-induced interactions in materials of medium density and composition.

We suggest that it is possible that the neutron multiplicity in the muon-induced nuclear cascades and electromagnetic cascades is responsible for this difference. The experimental results from Refs.~\cite{dongming,lbb} show a higher neutron multiplicity than that predicted by FLUKA and we propose a neutron multiplicity correction function to correct the
neutron production rate.  This function is obtained by extrapolating the variation in neutron multiplicity as a function of muon energy between the proposed parameterization based 
on the measurements~\cite{dongming,lbb} and the FLUKA simulation. 
\begin{equation}
\label{eq:correction}
\frac{M_{d} - M_{mc}}{M_{d}} = 0.64E_{\mu}^{0.02} - 0.74E_{\mu}^{-0.12},
\end{equation}
where $M_{d}$ is the measured neutron  multiplicity, $M_{mc}$ is the simulated
neutron multiplicity in FLUKA and $E_{\mu}$ is the muon energy in GeV. 
After correcting the neutron multiplicity in the FLUKA simulation, good agreement is found
between the data and the simulation as can be seen in Fig.~\ref{fig:agreement}.
\begin{figure}[htb!!!]
\includegraphics[angle=0,width=8.6cm]{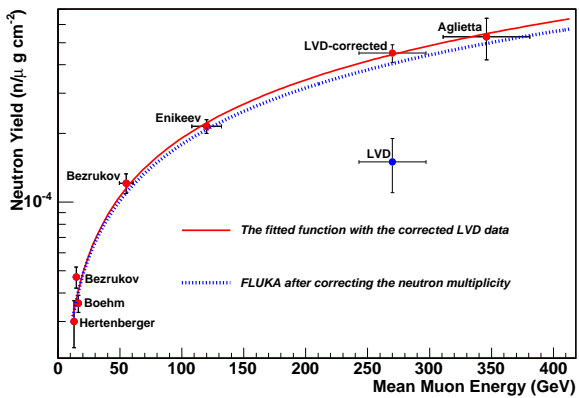}
\caption{\small{
The muon-induced neutron production rate versus the mean muon energy after correcting the
neutron multiplicity in the FLUKA simulation.}}
\label{fig:agreement}
\end{figure}

Further improvement might be gained with minor modifications in the inclusion of deep inelastic scattering of muons on nucleons. More generally, it is desirable to have more data for high energy muon interactions in the appropriate materials in order to more accurately tune the simulations relevant to neutron production deep underground. Nonetheless, whether we use our global fit function to the measured data or rely upon the multiplicity-corrected FLUKA simulation, the muon-induced neutron yield reproduces the data within an accuracy of about 15\%.

\subsection{\label{scuts}Media Dependence of Neutron Production Rate}
The muon-induced production rate for neutrons depends critically on knowledge of the chemical composition and density of the medium through which the muons interact.  We have studied this dependence using the FLUKA simulation specific to Gran Sasso in order to compare directly with Ref.~\cite{afk}. The dependence on atomic weight is shown in Fig.~\ref{fig:a}, where the general trend is well described by a power law, consistent with Ref.~\cite{afk} using slightly different fitting parameters.
\begin{equation}
\label{eq:atomic1}
<n> = 4.54 \times 10^{-5} {A}^{0.81} n/(\mu\; g\; cm^{-2}).
\end{equation}
         
\begin{figure}[htb!!!]
\includegraphics[angle=0,width=8.6cm]{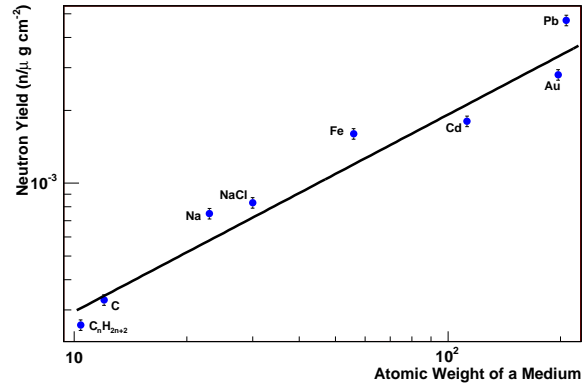}
\caption{\small{
Simulation of the muon-induced neutron production rate versus the atomic weight of the medium.}}
\label{fig:a}
\end{figure} 
   
The contribution to the neutron production rate from electromagnetic showers becomes more important for a heavy target, since the cross-section of an electromagnetic muon interaction is 
proportional to Z$^{2}/A$. Fig.~\ref{fig:neutronrate} shows this dependence where, again, the general trend can be described using a power law:
\begin{equation}
\label{eq:atomic}
<n> = 1.27 \times 10^{-4} (\frac{Z^{2}}{A})^{0.92} n/(\mu\; g\; cm^{-2}).
\end{equation}
 
\begin{figure}[htb!!!]
\includegraphics[angle=0,width=8.6cm]{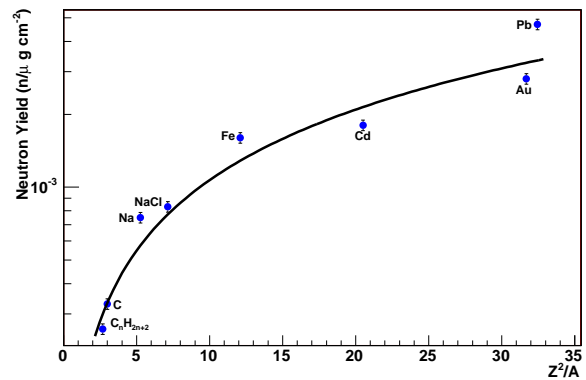}
\caption{\small{
Simulation of the muon-induced neutron production rate versus Z$^{2}/A$ of the medium.}}
\label{fig:neutronrate}
\end{figure}

\subsection{Neutron Fluxes and Differential Spectra at Underground Sites}
\subsubsection{Neutron Flux at Rock/Cavern Boundary}

The muon-induced neutron flux emerging from the rock into the cavern has been estimated
for the various underground sites considered in this work. We derive the neutron flux utilizing 
the FLUKA simulation with the corrected neutron multiplicity (equation~(\ref{eq:correction})) and the muon fluxes and distributions outlined in Section II. The neutron flux
($\phi_{n}$) as a function of depth is shown in Fig.~\ref{fig:neutronflux} where we have included a fit function of the following form:
\begin{equation}
\label{eq:neutronfluxtotal1}
\phi_{n} = P_{0}(\frac{P_{1}}{h_{0}})e^{-h_{0}/P_{1}},
\end{equation}
where h$_{0}$ is the equivalent vertical depth (in km.w.e.) relative to a flat overburden. The fit parameters are
P$_{0}$ = (4.0 $\pm$ 1.1)$\times10^{-7}$ cm$^{-2}$s$^{-1}$ and P$_{1}$ = 0.86 $\pm$ 0.05 km.w.e..
 
\begin{figure}[htb!!!]
\includegraphics[angle=0,width=8.6cm]{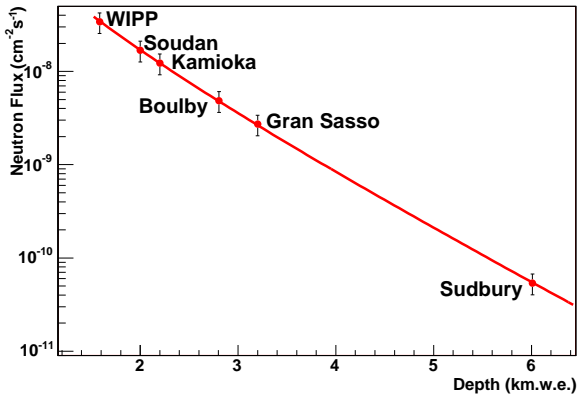}
\caption{\small{
The total muon-induced neutron flux deduced for the various underground sites displayed. Uncertainties on each point reflect those added in quadrature from uncertainties in knowledge of the absolute muon fluxes and neutron production rates based upon our simulations constrained by the available experimental data.}}
\label{fig:neutronflux}
\end{figure} 

In Table~\ref{tab:tabneu1} we summarize the neutron flux at the rock/cavern boundary
for the various sites considered and note that we have not included the effect of neutrons that emerge from one surface and back-scatter back into the cavity. The results are in good agreement
with the existing simulation results for Gran Sasso~\cite{ade}. If the simulation
results for Boulby~\cite{mjc} are modified using our neutron multiplicity correction,
 good agreement is also found between the two results. It is relevant to note that there is a significant fraction of the neutrons with energy above 10 MeV.

\begin{table}[htb!!!]
\caption{The muon-induced neutron flux for six sites (in units of 10$^{-9}$ cm$^{-2} s^{-1}$). The total flux is included along with those predicted for neutron energies above 1, 10, and 100 MeV.}
\label{tab:tabneu1}
\begin{tabular}{l|l|l|l|ll}
\hline \hline
Site & total&$>$ 1.0 MeV & $>10 MeV$ & $>100 MeV$\\
\hline
WIPP &34.1&10.78 & 7.51&1.557\\
Soudan &16.9&5.84 & 4.73&1.073\\
Kamioka &12.3&3.82 & 3.24&0.813\\
Boulby &4.86&1.34&1.11&0.277\\
Gran Sasso &2.72&0.81&0.73&0.201\\
Sudbury &0.054& 0.020&0.018&0.005\\
\hline \hline
\end{tabular}
\end{table}
 
\subsubsection{Neutron Production in Common Shielding Materials}
Fast neutrons can also be created by muons passing through the materials commonly used to shield a detector target from natural radioactivity local to the surrounding cavern rock. Fig.~\ref{fig:fluxshielding} shows the neutron yield in some common shielding materials. We have also included a simulation for germanium which will prove useful later in this paper when we consider the DSR for experiments based on this target material.

\begin{figure}[htb!!!]
\includegraphics[angle=0,width=8.6cm]{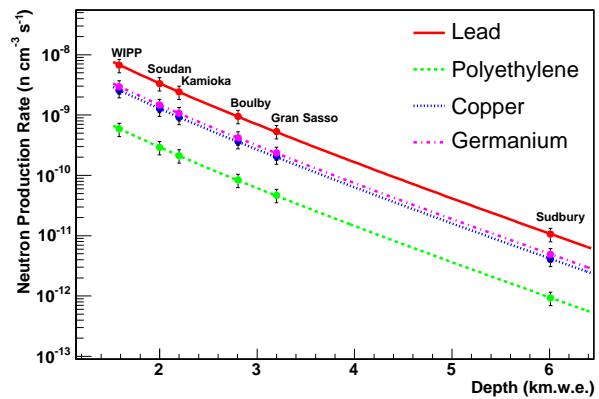}
\caption{\small {The muon-induced neutron production rate predicted for some common detector shielding materials. Note that minor variations due to neutron back-scattering have been neglected in these calculations.}}
\label{fig:fluxshielding}
\end{figure}
   
The fitted functions have the same form as equation~(\ref{eq:neutronfluxtotal1}) but with different values for parameters which are provided in 
Table~\ref{tab:parameters}. To convert the neutron production rate to the total neutron flux, one multiplies equation~(\ref{eq:neutronfluxtotal1}) 
by the average muon path length which depends upon the detector
geometry.

\begin{table}[htb!!!]
\caption{Summary of the fitting parameters describing the muon-induced neutron production rate in common detector shielding materials.}
\label{tab:parameters}
\begin{tabular}{lll}
\hline \hline
Material & P$_{0}$ & P$_{1}$ \\
\hline
Lead &(7.84$\pm2.21)\times10^{-8}$&0.86$\pm$0.05 \\
Polyethylene &(6.89$\pm1.95)\times10^{-9}$&0.86$\pm$0.05 \\
Copper &(2.97$\pm0.838)\times10^{-8}$&0.87$\pm$0.05\\
Germanium &(3.35$\pm0.95)\times10^{-8}$&0.87$\pm$0.05 \\
\hline \hline
\end{tabular}
\end{table}

Generally speaking, muon-induced neutrons produced in a detector target or surrounding shield can be actively vetoed in coincidence with the primary, depending on the veto efficiency and specific detector geometry. Specific examples are provided later in this paper.

\subsubsection{Neutron Energy Spectrum}
It is well known that the energy spectrum of neutrons produced by muon spallation is
uncertain~\cite{mal, yfg, afk, jcb, vch} and that data is scarce, particularly for studies relevant for deep underground sites. Nonetheless, with reference to Fig.~\ref{fig:energyspectrum} and our previous discussion of the LVD data sample obtained at Gran Sasso, the FLUKA simulations reproduce the data well once the data are appropriately corrected for the quenching of proton-recoil energy. Recently, Ref.~\cite{vch} reported a measurement
of the muon-induced neutron energy spectrum using 190 GeV/c muon interactions
on a graphite target. The neutrons were observed by liquid scintillator detectors and the
neutron energy distribution was determined via time-of-flight. The measured angular and energy distributions agree well with the FLUKA simulation performed by Ref.~\cite{yfg}.
 
\begin{figure}[htb!!!]
\includegraphics[angle=0,width=8.6cm]{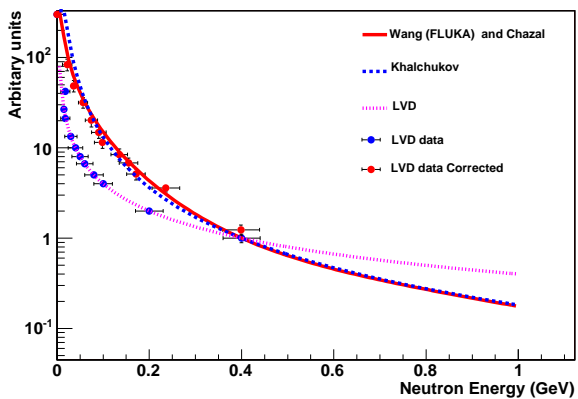}
\caption{\small {The differential energy spectrum for muon-induced neutrons as measured in the LVD experiment before and after correcting for proton-recoil quenching effects described in the text. Following such corrections, the FLUKA simulation appears to reproduce the shape of the spectrum well.}}
\label{fig:energyspectrum}
\end{figure}
 
We derive the neutron energy spectrum for each experimental site (Fig.~\ref{fig:neuspe}) from the FLUKA simulation for the neutrons produced in the rock and then emerge into the experimental hall. 
The muons are generated locally for each site as described in Section II and used as input to the FLUKA simulation.

\begin{figure}[htb!!!]
\includegraphics[angle=0,width=8.6cm]{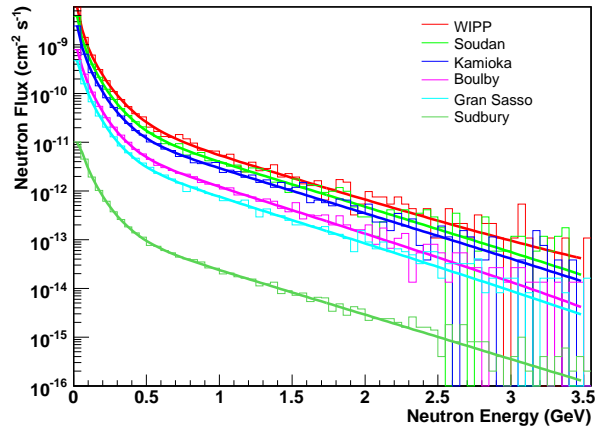}
\caption{\small {The differential energy spectrum for muon-induced neutrons at the various underground sites. The bin width is 50 MeV.}}
\label{fig:neuspe}
\end{figure}

For each site we provide a convenient parameterization based upon the following fitting function:
\begin{equation}
\label{eq:neuspe}
\frac{dN}{dE_{n}} = A_{\mu}(\frac{e^{-a_{0}E_{n}}}{E_{n}}+B_{\mu}(E_{\mu})e^{-a_{1}E_{n}})+a_{2}E_{n}^{-a_{3}},
\end{equation}  
where A$_{\mu}$ is a normalization constant, a$_{0}$, a$_{1}$,  a$_{2}$  and a$_{3}$ are
fitted parameters, E$_{n}$ is the neutron energy, B$_{\mu}$(E$_{\mu}$) is a function of muon energy and E$_{\mu}$ is in GeV,
\begin{equation}
\label{eq:bmuon}
B_{\mu}(E_{\mu}) = 0.324 - 0.641e^{-0.014E_{\mu}}.
\end{equation} 
This parameterization is consistent with Ref.~\cite{yfg} and is valid for E$_{n}>$10 MeV. 
The fit parameters and the average neutron energy for each site are summarized in Table~\ref{tab:neutab}.

\begin{table}[htb!!!]
\caption{Summary of the fitting parameters describing the shape of the differential energy spectrum of muon-induced neutrons for the various underground sites.}
\label{tab:neutab}
\begin{tabular}{lllllll}
\hline \hline
Site & $<E>$ & a$_{0}$ & a$_{1}$ & a$_{2}$ & a$_{3}$ \\
\hline
WIPP  & 62 MeV & 6.86 & 2.1&2.971$\times10^{-13}$&2.456\\
Soudan& 76 MeV& 7.333 & 2.105&-5.35$\times 10^{-15}$ & 2.893\\
Kamioka& 79 MeV & 7.55 & 2.118&-1.258$\times 10^{-14}$ & 2.761\\
Boulby & 88 MeV & 7.882 & 2.212 &-2.342$\times 10^{-14}$&2.613\\
Gran Sasso &91 MeV & 7.828 & 2.23&-7.505$\times10^{-15}$&2.831\\
Sudbury& 109 MeV & 7.774 & 2.134&-2.939$\times 10^{-16}$&2.859\\
\hline \hline
\end{tabular}
\end{table}

\subsubsection{Neutron Angular Distribution}
The angular distribution of neutrons produced in the rock by muons is shown in Fig.~\ref{fig:neangu}. As described in Refs.~\cite{yfg, kva}, our simulations reproduce the expected forward peak
for those neutrons that are produced largely through muon spallation whereas the secondary evaporation of neutrons is predominantly distributed isotropically along the muon track. 

\begin{figure}[htb!!!]
\includegraphics[angle=0,width=8.6cm]{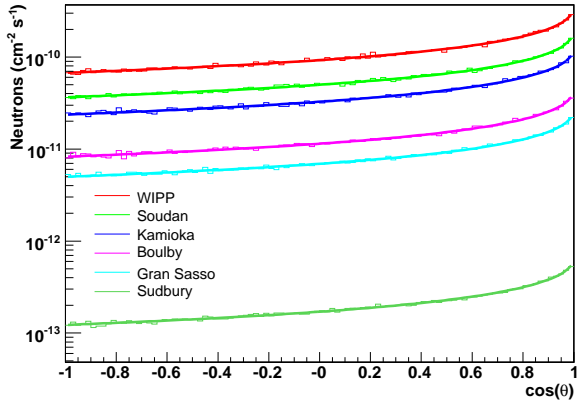}
\caption{\small {Simulation of the muon-induced neutron angular distribution for neutrons produced relative to the primary muon track.}}
\label{fig:neangu}
\end{figure}

We parameterize the angular distribution according to:
\begin{equation}
\label{eq:angular}
 \frac{dN}{dcos(\theta)} = \frac{A_{\theta}}{(1-cos(\theta))^{B_{\theta}(E_{\mu})}+C_{\theta}(E_{\mu})},
\end{equation}
where A$_{\theta}$ is a constant and B$_{\theta}$(E$_{\mu}$) and C$_{\theta}$(E$_{\mu}$) are 
weakly correlated to muon energy
and E$_{\mu}$ is in GeV.
The corresponding functions are: 
\begin{equation}
B_{\theta}(E_{\mu}) = 0.482E_{\mu}^{0.045}
\end{equation} 
and 
\begin{equation}
C_{\theta}(E_{\mu}) = 0.832E_{\mu}^{-0.152}.
\end{equation}

\subsubsection{Neutron Multiplicity}
The number of neutrons produced per muon interaction is the least known quantity
in the production of neutrons induced by muons. As discussed previously, the average multiplicity
in FLUKA is smaller than that of the measurements~\cite{dongming,lbb}. 
The neutron multiplicity distributions obtained from our simulations are shown in Fig.~\ref{fig:multi1}.
 
\begin{figure}[htb!!!]
\includegraphics[angle=0,width=8.6cm]{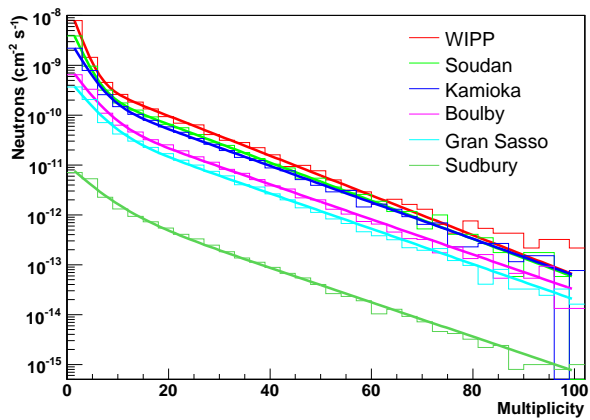}
\caption{\small{
Calculated neutron multiplicity at different experimental sites. Shown is a solely FLUKA calculation without correcting 
multiplicity using equation~(\ref{eq:correction}). 
}}
\label{fig:multi1}
\end{figure}
 
The parameterization function proposed by Ref.~\cite{yfg} is employed:
\begin{equation}
\label{eq:fun1}
\frac{dN}{dM} = A_{M}(e^{-B_{M}(E_{\mu})M}+C_{M}(E_{\mu})e^{-D_{M}(E_{\mu})M}),
\end{equation}
where A$_{M}$ is normalization constant, M is the multiplicity and E$_{\mu}$ is in GeV. 
We found
\begin{equation}
\label{eq:fun2}
B_{M}(E_{\mu}) = 0.321E_{\mu}^{-0.247},
\end{equation}
\begin{equation}
\label{eq:fun3}
C_{M}(E_{\mu}) = 318.1e^{-0.01421E_{\mu}},
\end{equation}
and
\begin{equation}
\label{eq:fun4}
D_{M}(E_{\mu}) = 2.02e^{-0.006959E_{\mu}}.
\end{equation}

The average multiplicity exhibits the expected dependence on muon
energy, and thus depth, and is apparent in the fit parameters $<M>$ = 3.48, 4.26, 5.17, 6.03, 6.44 and 7.86 for WIPP, Soudan, Kamioka, Boulby, Gran Sasso and Sudbury, respectively. The neutron multiplicity is also dependent on the different target materials. We have simulated the neutron distributions using the average density and chemical composition appropriate for each site and applied a correction to the simulated multiplicity according to equation~(\ref{eq:correction}). The corrected  multiplicity agrees with the result in Ref.~\cite{dongming} for KamLAND.
     
\subsubsection{Neutron Lateral Distribution}
 Fig.~\ref{fig:lateral} shows a FLUKA simulation of the lateral distribution of neutrons as they emerge from the primary muon track in a selection of media. Typically speaking, the neutron flux is attenuated by about two orders of magnitude at distances larger than ~3.5 m from the muon track, however, as much as ~10\% remain at distances as large as 2 to 2.5 m.
\begin{figure}[htb!!!]
\includegraphics[angle=0,width=8.6cm]{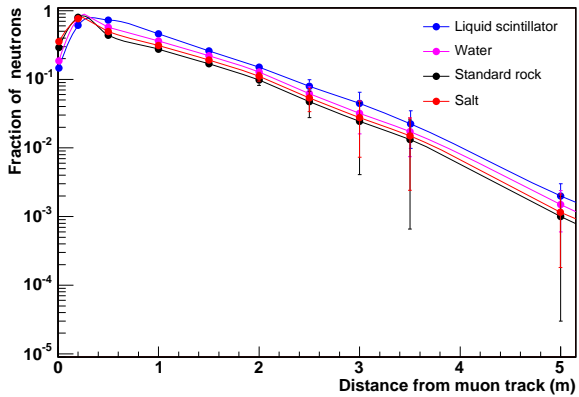}
\caption{\small{
The fraction of muon-induced neutrons emerging as a function of
distance from the primary muon track in several media. The curves exhibit distinct features relevant to neutron production and propagation in the media. At short range from the muon track, neutron production increases with distance as the nuclear and electromagnetic shower develops, however, neutron production weakens after about 50 cm from the muon track and propagation/attenuation of neutrons in the medium dominates.}}
\label{fig:lateral}
\end{figure}

\section{Muon-Induced Cosmogenic Activity}
As the cascades of muon-induced reactions propagate through detector materials, the
production rate of cosmogenic nuclide $j$ at depth $X$ in a detector volume can be expressed as: 
\begin{equation}
\label{eq:decay}
R_{j}(X) = \sum_{i} n_{i}\sum_{k}\int \sigma_{ijk}(E_{k}) \cdot \phi_{k}(E_{k},X)dE_{k},
\end{equation}
where n$_{i}$ is the number of atoms for target element $i$ per kg of material in the
detector, $\sigma_{ijk}(E_{k})$ is the cross section for the production of nuclide $j$ from the 
target element $i$ by particles of type $k$ with energy E$_{k}$, and  $\phi_{k}(E_{k},X)$ 
is the total flux of particles of type $k$ with energy E$_{k}$. The production cross sections are discussed in detail in Ref.~\cite{jso} where the equivalent photon approximation is used. The energy dependence of the corresponding cross-section can also be 
described by $\sigma_{\mu}$(E) = $\sigma_{0} \cdot$ E$_{\mu}^{0.7}$\cite{hei}.

Neutrons can interact with nuclei to produce long-lived radioactive
isotopes and secondary neutrons. The production rate can be estimated as
\begin{equation}
\label{eq:neu1}
R(h) = \int \Phi_{n}(E,h) \cdot N_{a} \cdot \sigma_{n}(E) dE,
\end{equation}
where  $\Phi_{n}(E,h)$ is the flux of neutrons on the detector at  depth $h$, N$_{a}$ is the number of atoms of the target
 and  $\sigma_{n}$(E) 
represents the
production cross section of the neutron reaction~\cite{tan,fuk,fua}. 
The neutron flux $\Phi_{n}$(E) depends on the neutron energy. 
Note that the long-lived radioactive isotopes produced by neutrons near the Earth's 
surface are the dominant product of the muon-induced cosmogenic radioactivity. 

The production of cosmogenic radioactivity depends strongly on the target and it must be 
evaluated specifically for an individual experiment. Nonetheless, the production rate is proportional to the muon, or neutron, flux and the relevant interaction cross-section.  The energy dependence of the total cross-section for all muon-induced radio-isotopes in liquid scintillator
was evaluated assuming the power law~\cite{ffk},
\begin{equation}
\label{eq:cross1}
\sigma_{tot}(E_{\mu}) \propto E_{\mu}^{\alpha},
\end{equation}
where $\alpha$ varies from 0.50 to 0.93 with a weight mean value $<\alpha>$ = 0.73$\pm$0.10
~\cite{hag}. For a target of N atoms and cross-section $\sigma_{0}$ at the Earth's
surface, where the average muon energy is about 4 GeV, the muon-induced cosmogenic radioactivity (R$_{iso}$) depends on the differential muon energy spectrum dN$_{\mu}$/dE$_{\mu}$ at the experimental site:
\begin{equation}
\label{eq:cross2}
R_{iso} = N\sigma_{0}\int_{0}^{\infty} (\frac{E_{\mu}}{1\; GeV})^{\alpha} \frac{dN_{\mu}}{dE_{\mu}}dE_{\mu},
\end{equation}
As a simplification, the production rate is written as a function of the average
muon energy $<E_{\mu}>$~\cite{hag}:
\begin{equation}
\label{eq:cross3}
R_{iso} = \beta_{\alpha}N\sigma_{1\; GeV}(\frac{<E_{\mu}>}{1\; GeV})^{\alpha}\phi_{\mu},
\end{equation}
where $\phi_{\mu}$ is the total muon flux at the experimental site and $\beta_{0.73}$ = 0.87 $\pm$ 0.03
is the correction factor for the averaging of E$_{\mu}$~\cite{hag}. For a given detector target,
a simple scaling relation, or Depth-Sensitivity-Relation (DSR) factor $F$, can be derived,
\begin{equation}
\label{eq:cross4}
F \equiv \frac{R_{iso}(Surface)}{R_{iso}(Underground)} = (\frac{4\; GeV}{<E_{\mu}>})^{\alpha}\frac{\phi_{\mu}(Surface)}{\phi_{\mu}(Underground)},
\end{equation}
which describes the reduction in muon-induced activity as one moves to deeper and deeper sites.
Table~\ref{tab:sensitivity} summarizes this effect.
\begin{table}[htb!!!]
\caption{The scaling factor, $F$, relevant to the Depth-Sensitivity-Relation (DSR) developed for the underground sites considered in this work.}
\label{tab:sensitivity}
\begin{tabular}{llll}
\hline \hline
Site &Depth & $<E_{\mu}>$ &  F \\
     &(km.w.e.)&(GeV) &   \\
\hline
WIPP  & 1.585  & 184  & 2140\\
Soudan& 1.95  & 212  & 4600\\
Kamioka& 2.05  & 219  & 5690\\
Boulby & 2.805 & 264  & 19180\\
Gran Sasso & 3.1 & 278  &29270\\
Sudbury& 6.011  & 356  & 1.67$\times10^{6}$\\
\hline \hline
\end{tabular}
\end{table}

\section{The Depth-Sensitivity-Relation}
In this section we develop the Depth-Sensitivity-Relation (DSR) for the major components of the muon-induced background, namely the cosmic-ray muons themselves, the induced neutron background, and cosmogenic radioactivity. In Fig.~\ref{fig:sensall} we show a global view for the DSR where we have arbitrarily normalized the DSR Factor $F$ at the shallower depth characteristic of the WIPP site. Generally speaking, the muon flux and induced activity is reduced by about one order of magnitude for every increase in depth of 1.5 km.w.e.. 

\begin{figure}[htb!!!]
\includegraphics[angle=0,width=8.6cm]{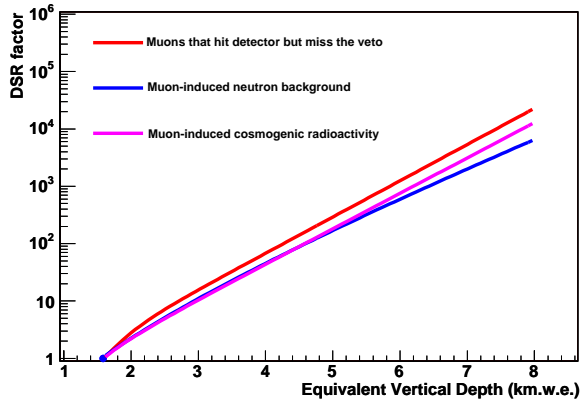}
\caption{\small{
Relative attenuation factors for the muon and muon-induced activities as a function of overburden. The curves are normalized, arbitrarily, to unity for the shallower depth defined by the WIPP site at 1.585 km.w.e.. Roughly, an attenuation factor of about one order of magnitude is achieved for every 1.5 km.w.e. increase in depth.}}
\label{fig:sensall}
\end{figure}

The curves shown in Fig.~\ref{fig:sensall} are indicative of the relative muon flux and muon-induced activity that will be present for a given laboratory site at its characteristic depth. The effect of this activity will depend on the specific details of a given detector geometry, including shielding, and the goals of a particular experiment. Generally speaking, muons that traverse a detector unvetoed and muon-induced fast neutrons are the primary concern deep underground, while long-lived cosmogenic activity is usually dominated by activation of detector materials at the surface and prior to construction underground. In what follows, we describe simulations for a germanium-based detector and apply this to develop the DSR for specific examples in the search for dark matter and neutrinoless double beta decay. Particular attention is paid to the sensitivity to muon-induced fast neutrons.

\subsection{Simulation Set-Up for Germanium Based Detectors}

To evaluate the response of neutrons in a detector, a Monte Carlo 
simulation code  has been developed to simulate the neutrons generated in different media where we rely on the neutron fluxes and distributions generated and discussed above. The detector geometry, material, and electromagnetic interactions are simulated using GEANT 3~\cite{rbr}. Hadronic interactions are simulated using the Nucleon Meson 
Transport Code, NMTC~\cite{czt}, while transportation of low energy neutrons is achieved using GCALOR~\cite{czt}. Fast neutrons deposit their energy via elastic scattering and/or inelastic scattering processes. We will demonstrate that the former is the main concern for dark matter searches since the elastic scattering process tends to deposit energy in the low energy region of interest while the latter dominate the background through inelastic scattering process owing to the ensuing $\gamma$-rays produced above the Q-value for double beta decay.
In general, inelastic reactions of fast neutrons leave the residual nucleus in a highly excited state which subsequently decays via $\gamma$-cascades to the ground state in typically three or four steps. The initial intensity distribution over a very large number
of highly excited levels is collected in the first few excited levels which then decay to the ground state.
In the simulation, the Hauser-Feshback theory~\cite{wha} is used to calculate inelastic scattering cross sections for excitation of a given level depending on the
properties of the ground state and the excited state. This theory was first formed by Hauser and Feshback in the 1950's and later modified by
Moldauer~\cite{pam}. Since then many experiments have verified the theory~\cite{hvo,ric,apa}. 

In addition to the geometry associated with the detector and shielding materials, it is important also to define the geometry and dimension of the cavern housing the experiment.
For example, we have demonstrated that the neutron flux incident on the shielding around 
a detector can vary by factors of about 2-3, depending on the cavern size, due to
the back scattering of neutrons from the cavern walls. As such we specify a cavern size 30 $\times$ 6.5 $\times 4.5 \;m^{3}$ in our simulations. The effects of lead, polyethylene, copper, and target material on neutron production
and absorption are also important to the neutron simulation. We have seen a large increase 
(a factor $\sim$10-20, depending on the thickness of lead)
in the neutron flux due to the additional and efficient production of neutrons in lead, an effect that has also been identified in Ref.~\cite{cbu}. Consequently, the DSR developed in what follows should be understood within the boundary conditions described for the specific experiments considered.
      
\subsection{DSR for Dark Matter Experiments}
Experiments geared toward the direct detection of dark matter such as WIMPs (Weakly Interacting Massive Particles) rely on detector technologies capable of visible energy thresholds well below ~100 keV in order to observe the recoil energy induced via WIMP scattering off the nucleus. In order to have sufficient sensitivity to the feeble WIMP cross-section, such detectors must also be constructed of materials with extremely low levels of natural radioactivity and be able to discriminate background from ionizing $\gamma$-rays and electrons that can otherwise fog a potential WIMP signal. With this discrimination power in hand, it remains to assure that nuclear recoil events associated with fast neutrons are kept sufficiently rare as they present an ineluctable background in the search for WIMPs. To date, the most stringent limits on the WIMP-nucleon cross-section ($\sim 1.6 \times 10^{-43}$$ cm^2$) have been provided by the CDMS-II experiment operating in the Soudan mine~\cite{cdms} and it is the goal of next generation experiments to improve this sensitivity by several orders of magnitude.

In order to demonstrate the sensitivity of dark matter experiments to muon-induced fast neutrons we derive the DSR for the CDMS-II detector~\cite{cdms2}, which consists of a tower of four Ge (250 g) and two Si (100 g) detectors surrounded by an average of
0.5 cm of copper, 22.5 cm of lead and 48.6 cm of polyethylene. A 5-cm-thick muon veto detector with efficiency $>$ 99.9\% encloses the shielding.
 
The production rate (R) of nuclear recoil events produced by fast neutrons can be expressed as:
\begin{equation}
\label{eq:neucdms}
R_{i}(h) =  \sum_{i} n_{i} \int \frac{d\Phi_{n}(E_{n})}{dE_{n}} \sigma_{i}(E_{n}) F_{i}(E_{n}) dE_{n},
\end{equation}
where $n_{i}$ is the number of atoms for target element $i$ per $kg$ material in the
detector, $\frac{d\Phi_{n}(E)}{dE}$ is the neutron energy spectrum (equation~(\ref{eq:neuspe})) at depth $h$,
$\sigma_{i}(E)$ is the neutron interaction cross section~\cite{tan,fuk,fua} with $i^{th}$ element of natural Ge, and
 $F_{i}(E)$ is an energy-dependent quenching function~\cite{lind} specific to the $i^{th}$ element of Ge. 

We generate the muon-induced neutrons at the rock/carven boundary using the formalism outlined in Section III and propagate them through the CDMS II geometry described above. Since the muon veto in CDMS II has an efficiency greater than 99.9\%~\cite{kama}, 
we are concerned only for the neutrons produced in the rock. We have performed our simulations for CDMS II using two different shielding configurations. Shielding
configuration 1 is that used in the actual experiment with 0.5 cm copper, followed by 8.6 cm polyethylene, 22.5 cm lead, and 40 cm of polyethylene as the outer neutron shield. In shielding configuration 2, we interchanged the thick polyethylene and lead shield positions. In this case we found a reduction in background by about a factor of two over the CDMS-II shield. This reduction occurs owing to the additional neutrons produced when neutrons from the rock interact in the lead shield. Similar observations have been made in Ref.~\cite{cbu} and Ref.~\cite{mjc}. The visible recoil energy spectrum induced by the fast neutrons is shown in Fig.~\ref{fig:wimp} for three different depths and along with the spectrum expected for dark matter assuming a cross section $\sigma_{p}$ = 10$^{-46} cm^{2}$ and a 100 GeV WIMP mass.

\begin{figure}[htb!!!]
\includegraphics[angle=0,width=8.6cm]{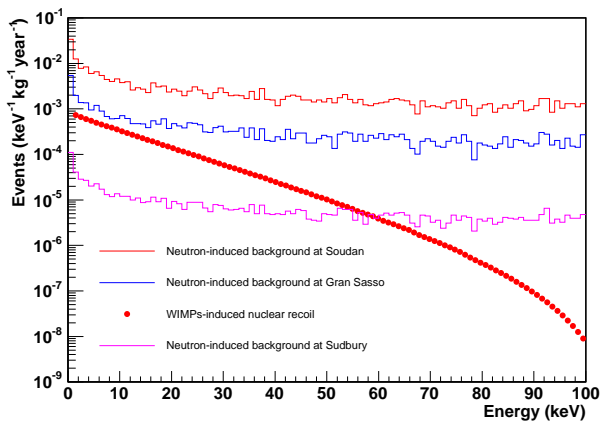}
\caption{\small{The predicted event rates for spin-independent WIMP-nucleon scattering (dotted-line) in Ge assuming a WIMP-nucleon cross-section of
$\sigma_{p}$ = 10$^{-46} cm^{2}$ and a 100 GeV WIMP mass. Muon-induced neutron backgrounds are also displayed for comparison, indicating the need for greater and greater depth as experiments evolve in scale and sensitivity.}}
\label{fig:wimp}
\end{figure}

Using these results we determine an event rate of 0.9 events/kg-year in an energy window of 10 to 100 keV for the CDMS-II experiment operating at the Soudan mine. This rate is reduced to 0.5 events/kg-year after identifying those neutrons that interact with two or more crystals in the CDMS-II tower. Our prediction is consistent with the upper bound of 0.94 $\pm$ 0.38  events/kg-year that can be deduced from the CDMS II collaboration's limit of 
0.05 $\pm$ 0.02  neutrons detected during an exposure of 19.4 kg-day~\cite{cdms2} or 34 kg-day~\cite{cdms}. In Ref.~\cite{kama}, Kamat also simulated the un-vetoed neutron rate in the CDMS-II detector and obtained 0.05 $\pm$ 0.02 neutrons  for the 19.4 kg-day exposure, in excellent agreement with our prediction.

We can now derive the DSR appropriate to the CDMS-II experiment. As shown in Fig.~\ref{fig:sensdepth}, the experiment's sensitivity would be limited to $\sigma_{p}\sim 10^{-44} cm^{2}$ due to the muon-induced fast neutron flux at Soudan and that depths in excess of $\sim$ 5 km.w.e. will be required to push beyond $\sigma_{p}\sim 10^{-46} cm^{2}$, unless the neutron flux can be suppressed effectively either by further shielding and/or active veto.

\begin{figure}[htb!!!]
\includegraphics[angle=0,width=8.6cm]{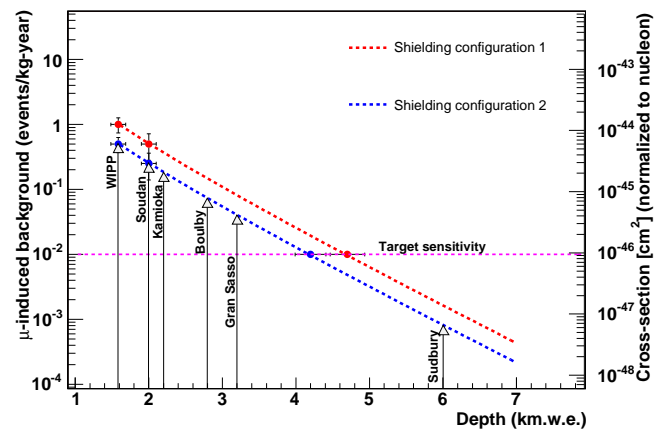}
\caption{\small{The Depth-Sensitivity-Relation (DSR) derived for the CDMS-II detector geometry for the two shielding configurations described in the text. The muon-induced background is dominated by elastic scattering of neutrons depositing visible energy in a 10 to 100 keV window. Specific points are shown, for example, at the depth of the Soudan mine where the CDMS-II detector has been operating. Uncertainties reflect those present due to uncertainties in the rock composition and in generating the muon-induced fast neutron flux.}}
\label{fig:sensdepth}
\end{figure}

It was pointed out in Ref.~\cite{mjc} that the
nuclear recoil event rate in coincidence with a second energy
deposition not associated with nuclear recoils (electrons, photons, muons etc.)
is a factor of 10 more than the rate of isolated nuclear recoil. We cannot confirm
this statement (a factor of 10) with the CDMS II geometry for the neutrons
which are produced in the rock and associated muons that miss the veto.
This is likely due to the fact that the CDMS II detectors are segmented and much smaller than that considered in Ref.~\cite{mjc}. In our simulation, heavy charged particles such as
pions, kaons and protons with kinetic energy greater than 10 MeV that are produced together
 with neutrons by muons 
in the rock
are a factor of 10 less
than that of neutrons and they are about 40 times smaller 
in number than
 the neutrons after passing through the shielding. The
electrons and bremsstrahlung photons that are produced in the rock cannot survive the rock and air.
 The bremsstrahlung photons that are produced in the lead through nuclear showers induced by 
neutrons could generate the recoils in the detector. However, this contribution is limited by 
the size of the detectors and most of these events are multiple crystal events. In the CDMS-II simulation considered here as an example, we find that only 10\% of the recoil energy deposited by fast neutrons are coincident with secondary particles that can potentially sum with the neutron energy deposited in the detector.

\subsection{DSR for Double Beta Decay Experiments}
To demonstrate the effect of muon-induced activity in the search for neutrinoless double beta decay, we consider the geometry proposed for the Majorana project~\cite{majorana} where a detector module is made up of 57, 1.05 kg, closely packed crystals of germanium enriched to 86 percent in $^{76}$Ge. While details of the shielding for Majorana are under development, we consider an inner-most layer with 10 cm copper, followed by 40 cm lead and an outer-most layer of 10 cm polyethylene. An active muon veto outside the passive shield is also assumed but one that is limited to 90\% efficiency to veto nucleons produced inside the shielding. 

Both elastic and inelastic reactions of muon-induced neutrons are considered, however, unlike the case for dark matter, the dominant source of muon-induced background for the Majorana geometry results from the high-energy cascades that evolve from inelastic neutron scattering on the detector and shielding materials that produce background in the region of interest of the Q-value at 2039 keV. The results of our simulation are shown in Fig.~\ref{fig:back} with a breakdown of the main contributions summarized in Table~\ref{tab:contributions}. 

\begin{figure}[htb!!!]
\includegraphics[angle=0,width=8.6cm]{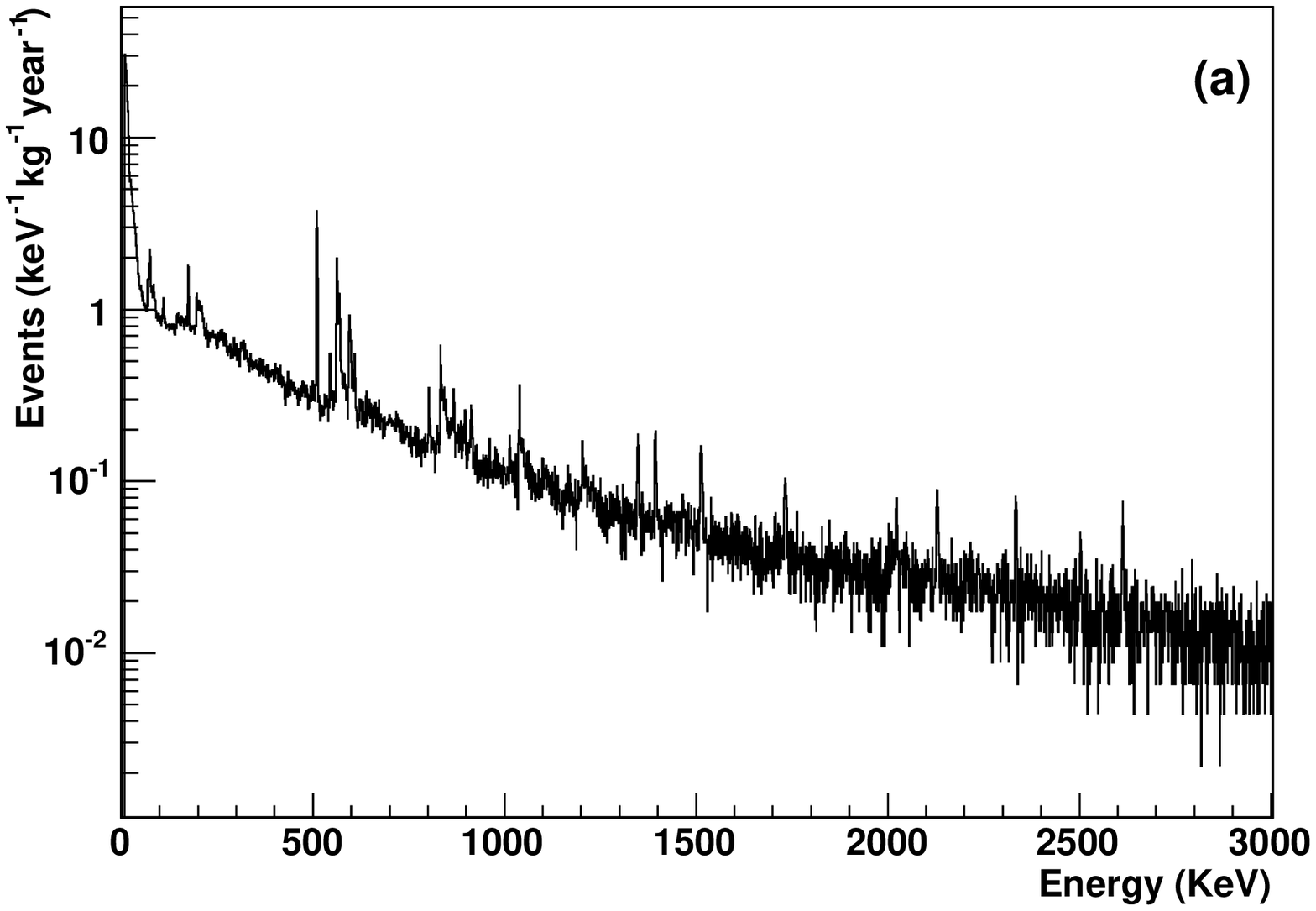}
\includegraphics[angle=0,width=8.6cm]{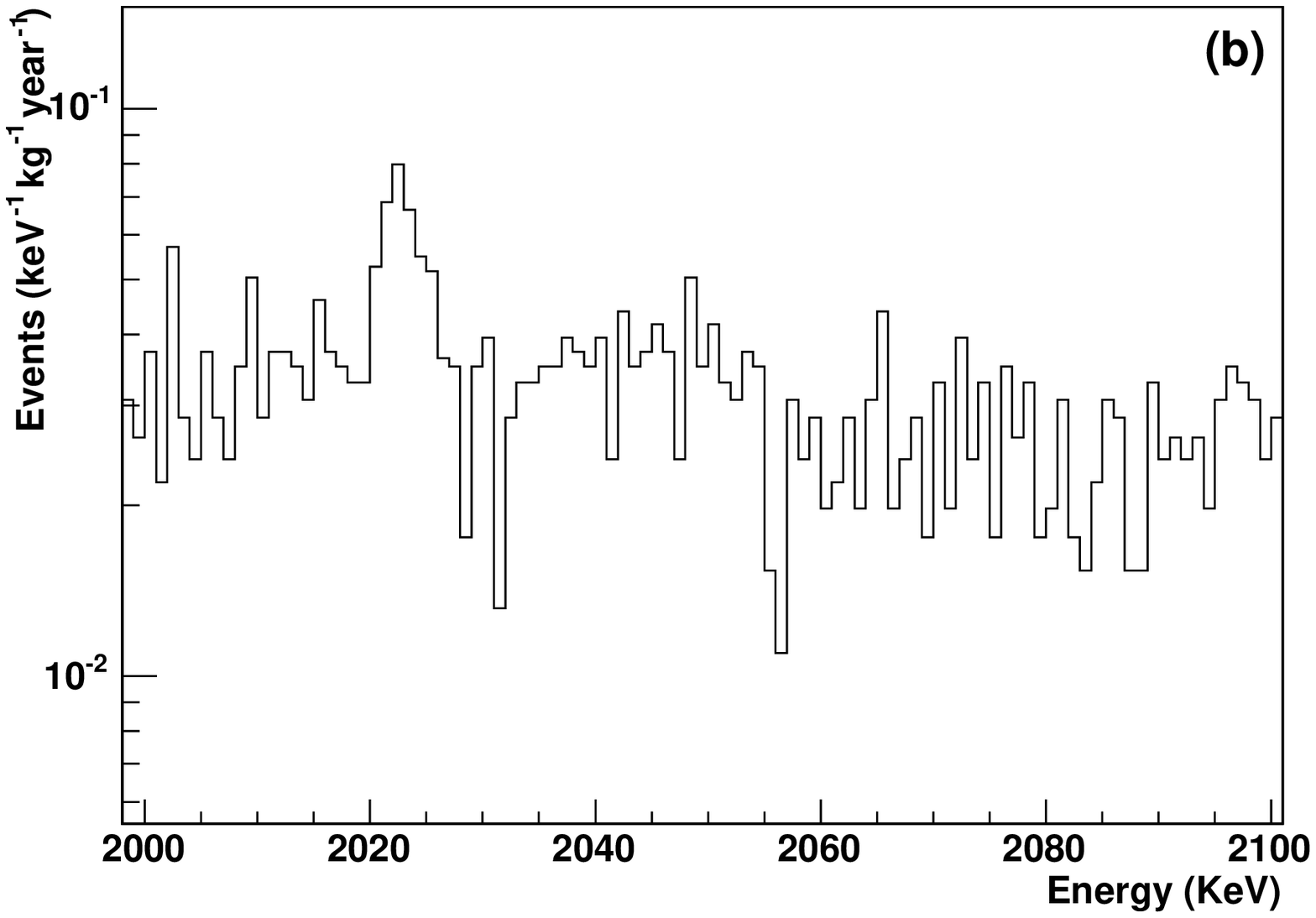}
\caption{\small{A simulation of the muon-induced background for a Majorana-like experiment operating at an equivalent overburden provided by the Gran Sasso Laboratory. In (a) we show the full spectrum with an expanded profile in (b) spanning the Region-of-Interest (ROI) around the Q-value for neutrinoless double beta decay at 2039 keV. The peak at 2023 keV is characteristic of that produced via the $^{76}Ge(n,n'\gamma)$ reaction.}}
\label{fig:back}
\end{figure}

\begin{table}[htb!!!]
\caption{Breakdown of the muon-induced background predicted for the energy range 2000-2100 keV in a Majorana-like experiment operating with an overburden characteristic of the Gran Sasso Laboratory.}
\label{tab:contributions}
\begin{tabular}{ll}
\hline \hline
Reaction & Events in the ROI \\
     &(events/keV-kg-year)  \\
\hline
 $^{76}Ge(n,n'\gamma$) & 0.01\\
 $^{74}Ge(n,n'\gamma$) & 0.002\\
 Cu$(n,n'\gamma$)  & 0.0019\\
 $^{208}Pb(n,n'\gamma)$&0.0035\\
 Elastic Scattering on Ge & 0.0036\\
 Muon hits & 0.0025\\
 Others& 0.0024\\
\hline \hline
\end{tabular}
\end{table}

Here we have performed our simulations assuming that the detector was operated at Gran Sasso depth in order to directly compare to previous germanium-based experiments situated there. In this case, we find that the total muon-induced background is about 0.026 events/keV-kg-year for Majorana at the depth of Gran Sasso. The dominant contribution (82\%) to this background results from inelastic neutron scattering processes ($Ge(n,n'\gamma)$, $Pb(n,n'\gamma)$ and $Cu(n,n'\gamma)$) on the detector target and shielding materials. Others (18\%) include stopping-muon capture on Ge, neutrons that capture on Ge and on Cu, and cosmogenic production in-situ.

It is interesting to set our simulations within the framework of the Heidelberg-Moscow experiment. Comparing our simulation to their background model~\cite{kdhk1}, we find agreement in the prediction of about 0.003 events/keV-kg-year due to events escaping the muon veto, however, we believe that the muon-induced neutron background and in situ cosmogenic production were missed in their simulation. We note that the simulated background in ref.~\cite{kdhk1} is about 20\% lower than that actually measured in the Heidelberg-Moscow experiment. Interestingly enough, this missing 20\%, corresponding to 0.022 events/keV-kg-year, is precisely what we have found in our simulation.

\begin{figure}[htb!!!]
\includegraphics[angle=0,width=8.6cm]{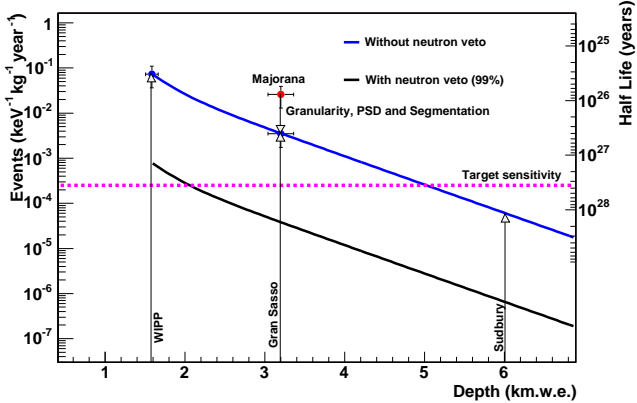}
\caption{\small{The Depth-Sensitivity-Relation (DSR) derived for a Majorana-like experiment showing, specifically, the results from this work assuming the detector is operated at a depth equivalent to the Gran Sasso Laboratory. The raw event rate in the energy region of interest of 0.026 events/keV-kg-year can be reduced by a factor of 7.4 by exploiting the detector granularity, pulse-shape discrimination (PSD), and detector segmentation. The upper curve displays the background simulated in the case that no active neutron veto is present and the lower curve indicates the reduction that would ensue if an active neutron veto were present that is 99\% efficient.}}
\label{fig:sensdepth1}
\end{figure}

The results of our simulations can be used to derive the DSR for Majorana as shown in Fig.~\ref{fig:sensdepth1}. The neutron induced background can be reduced by about a factor of 7.4 in Majorana owing to the use of crystal-to-crystal coincidences and the use of pulse-shape discrimination and segmentation. Nonetheless, to achieve the target sensitivity of next generation double-beta decay experiments,
0.00025 events/keV-kg-year corresponding to the background level required to reach sensitivity to the atmospheric mass scale of 45 meV Majorana neutrino mass, the muon-induced background must be reduced by roughly another factor of 100. This can be achieved only by operating such a detector at depths in excess of 5 km.w.e., otherwise an active neutron veto would need to be implemented with an efficiency in excess of 99\%.

\subsection{($\alpha$,n) Background}
Once the depth requirement is satisfied, a proper shield against ($\alpha$,n) neutrons from the
environment becomes necessary. We use the standard rock and 
the measured neutron flux (3.78$\times10^{-6} cm^{-2}s^{-1}$~\cite{belli,far} ) 
at Gran Sasso assuming that all underground labs have the same order of neutron flux
 to establish the shielding requirement for ($\alpha,n)$ neutrons. This flux
corresponds to an average of about 2.63 ppm $^{238}$U and 0.74 ppm $^{232}$Th activity in Gran Sasso rock and
 1.05 ppm $^{238}$U and 0.67 ppm $^{232}$Th activity in Gran Sasso concrete~\cite{hwu}. 
The neutron energy spectrum depending on the rock composition is shown in Fig.~\ref{fig:alphan}. 
As can be seen, the total neutron 
flux is about three orders of magnitude higher than that of neutrons from the rock due to muon-induced processes,
but the energy spectrum is much softer.
\begin{figure}[htb!!!]
\includegraphics[angle=0,width=8.6cm]{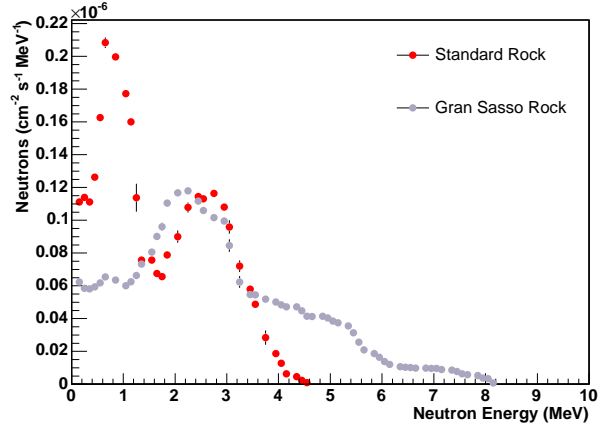}
\caption{\small{The neutron energy spectrum arising from ($\alpha$,n) reactions due to radioactivity in the rock. We predict a harder energy spectrum in Gran Sasso rock relative to standard rock owing to the presence of carbon and magnesium.}}
\label{fig:alphan}
\end{figure}

 To demonstrate the neutron flux and energy spectrum at different
boundaries we show the rock/cavern neutron flux and energy spectrum with 
a shielding for Majorana described earlier in Fig.~\ref{fig:bound}
 for the depth of Gran Sasso as an example.
\begin{figure}[htb!!!]
\includegraphics[angle=0,width=8.6cm]{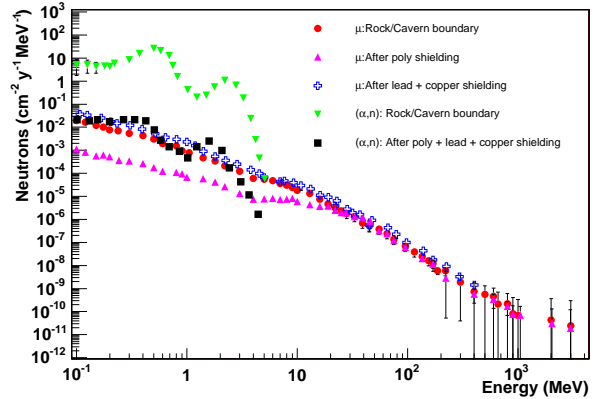}
\caption{\small{The energy spectrum for fast neutrons produced by ($\alpha,n)$ reactions in the rock compared to those induced by muon interactions in the rock with and without shielding. The lower energy neutrons ($<$ 10 MeV) are quickly absorbed using polyethylene shielding, however, the high energy portion of the muon-induced neutron flux persists. The addition of lead shielding adjacent to a detector can also create an additional source of muon-induced neutrons.}}
\label{fig:bound}
\end{figure}

Note that the ($\alpha,n)$ neutrons from the rock are quickly attenuated to the level of the muon-induced neutrons below ~10 MeV with rather moderate shielding whereas the higher energy muon-induced neutrons are essentially unaffected. We note also the increase in the muon-induced neutrons with the addition of lead shielding owing to the additional
neutron production in the heavy target. Consequently, the high energy muon-induced neutron background is the dominant concern given adequate shielding for the lower energy ($\alpha,n)$ neutrons. 

We show the shielding requirement for ($\alpha,n)$ neutrons  as 
a function of the thickness
of polyethylene in Fig.~\ref{fig:shield} 
in terms of the sensitivity of dark matter and double beta 
decay. Polyethylene shielding 30 to 40 cm thick is required for next generation experiments using Ge in the search for neutrinoless double beta decay while about 60 cm is required for dark matter searches.
\begin{figure}[htb!!!]
\includegraphics[angle=0,width=8.6cm]{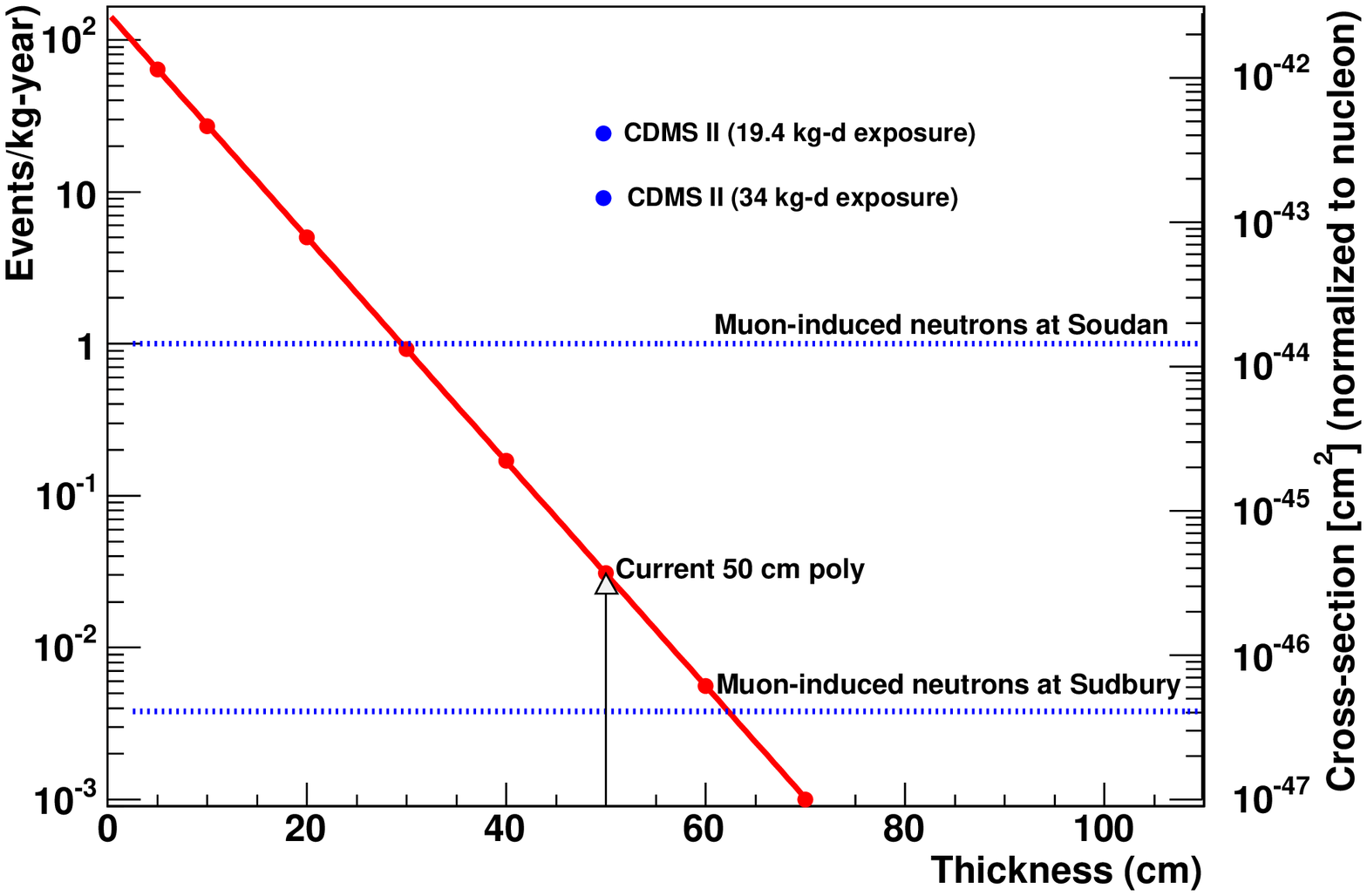}
\includegraphics[angle=0,width=8.6cm]{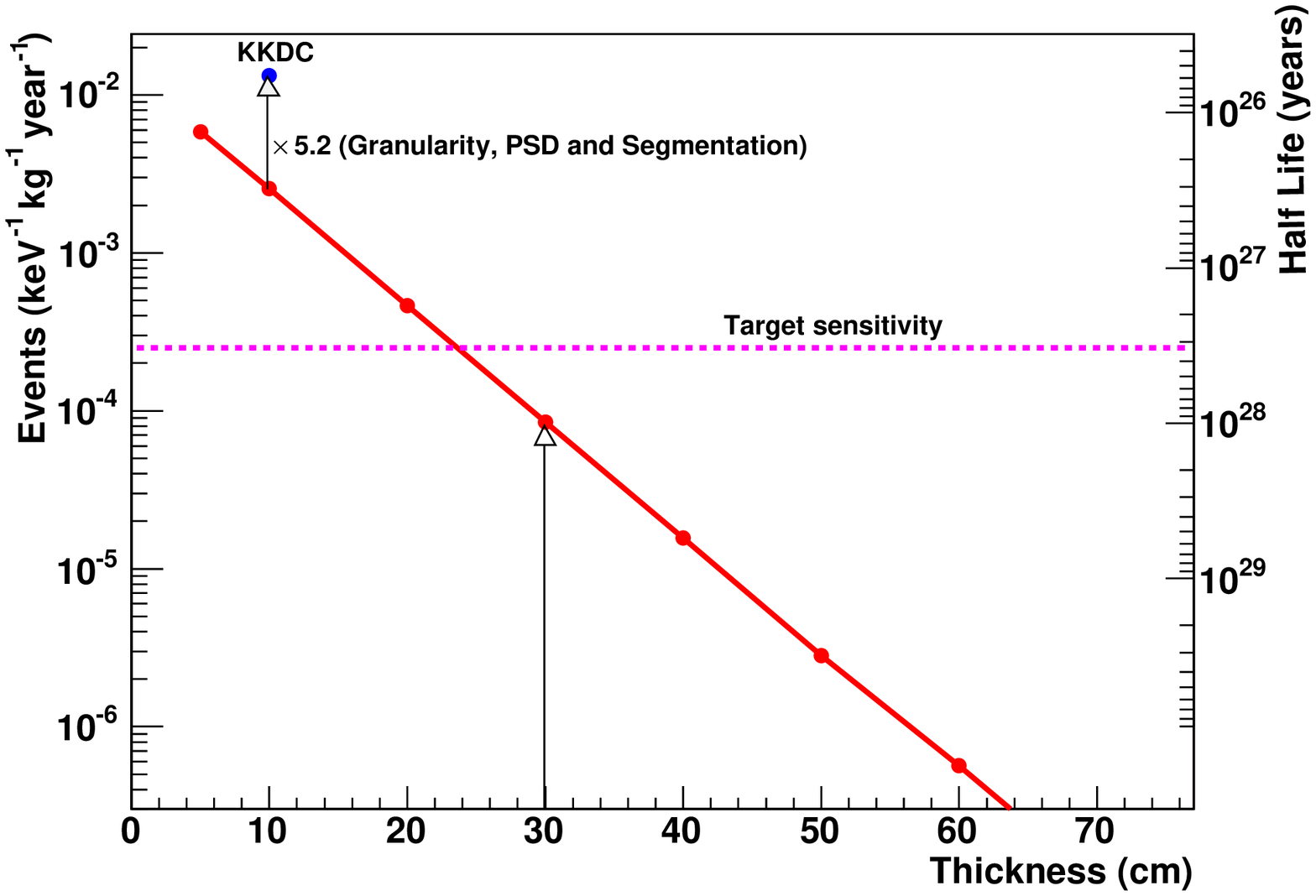}
\caption{\small{The ($\alpha$,n)-induced background versus polyethylene shielding thickness for the CDMS-II and Majorana detectors considered in this work. The upper limit on the spin-independent WIMP-nucleon cross section obtained by the CDMS II Collaboration~\cite{cdms} is shown in the upper panel for comparison along with that predicted for the muon-induced neutron background at Soudan and Sudbury. The lower panel includes our predicted value for the background in the Heidelberg-Moscow experiment (KKDC)~\cite{kdhk} before and after the reduction obtained using detector granularity, pulse-shape discrimination (PSD), and detector segmentation.}}
\label{fig:shield}
\end{figure}

\section{Summary}
We have provided a comprehensive study of the cosmic-ray muon flux and salient distributions as a function of depth and specific to a set of existing underground laboratories around the globe. We have applied these distributions to simulate the induced background at various underground sites and, where possible, made direct comparison to the available experimental data in order to assess the accuracy of our predictions. A Depth-Sensitivity-Relation has been developed and applied to examples of germanium-based detectors used in the search for cosmological dark matter and neutrinoless double beta decay.

The cosmic-ray muon flux is well described by a simple exponential law over a broad range in depth extending from about 1 to 8 km.w.e. We have defined depth in terms of the total muon flux obtained at an equivalent vertical depth to a site with flat overburden. This removes some of the confusion regarding the average depth often quoted for laboratories sited beneath mountains where the measured total muon flux is $\sim$ 15 to 20\% greater than what would be predicted based upon the average depth alone.

Good agreement can be found between the output of FLUKA simulations and the available experimental data on muon-induced fast neutrons provided one accepts our argument to correct the LVD data on both flux and energy distribution due to quenching effects. In that case we find that our simulations reproduce the data well, albeit with an overall normalization for the total neutron flux that appears to be underestimated by $\sim$ 35\%. This normalization appears to be greatly improved when one corrects the output of the FLUKA simulation to agree with experimental data on neutron multiplicity. Clearly, more data on the fast neutron energy spectrum and multiplicity induced by muons would be valuable to further bench-mark and tune the FLUKA simulations.

Our example DSR for dark matter searches is developed based on a model for the CDMS detector and demonstrates that depths in excess of $\sim$ 5 km.w.e. will be required in order to circumvent background from the elastic scattering of fast neutrons contaminating the low-energy region of interest for recoiling WIMPs. A similar conclusion can be made for neutrinoless double beta decay, modeled after the Majorana detector, where background following nuclear excitation due to the inelastic scattering of fast neutrons is the main culprit. Shallower depths make such experiments feasible provided the fast neutron flux can be adequately shielded and/or actively vetoed. The muon and muon-induced activity increases by approximately one order of magnitude for every decrease in depth of 1.5 km.w.e..

The program developed here has been applied to these specific types of experiments and detector geometries, however, the distributions presented in parameterized form can now be used as input to new simulations and background studies in other detectors of interest. The program could also be easily extended to underground sites under development that have not been considered in this work. More recently, we have begun an experimental program to verify some of our specific predictions by irradiating a Ge-detector with fast neutrons. Preliminary results indicate that the data agree well with our specific predictions for the Majorana detector. The details of that work is beyond the scope of this paper and will be communicated separately.

\section{Acknowledgments}
We are grateful to D. S. Akerib, J. Busenitz, S. R. Elliott, B. K. Fujikawa, V. A. Kudryavtsev, A. W. P. Poon, K. Rielage, R. G. H. Robertson, and J. F. Wilkerson for fruitful discussions. This work was supported in part by the U.S. Department of Energy and by the Los Alamos Directed Research and Development Program. 

%
%

\end{document}